\documentclass[prb,aps,twocolumn]{revtex4}
\usepackage{epsfig,epsf,psfrag}
\usepackage{graphicx}
\usepackage{epic,eepic}
\usepackage{color,pstcol}
\begin{document}
\title{Simple Floquet-Wannier-Stark-Andreev viewpoint and emergence
  of low-energy scales in a voltage-biased three-terminal Josephson
  junction}

\author{R\'egis M\'elin}
\affiliation{Centre National de la Recherche Scientifique, Institut
  NEEL, F-38042 Grenoble Cedex 9, France}

\affiliation{Universit\'e Grenoble-Alpes, Institut NEEL, F-38042
  Grenoble Cedex 9, France}

\author{Jean-Guy Caputo}

\affiliation{Laboratoire de Math\'ematiques, INSA de Rouen, Avenue de
  l'Universit\'e, F-76801 Saint-Etienne du Rouvray, France}

\author{Kang Yang}

\affiliation{Laboratoire de Physique Th\'eorique et des Hautes Energies,
CNRS UMR 7589, Universit\'e Pierre et Marie Curie,
Sorbonne Universit\'es, 4 Place Jussieu, 75252 Paris Cedex 05}

\affiliation{Laboratoire de Physique des Solides, CNRS UMR 8502,
  Univ. Paris-Sud, Universit\'e Paris-Saclay F-91405 Orsay Cedex, France}
\author{Beno\^{\i}t Dou\c{c}ot}

\affiliation{Laboratoire de Physique Th\'eorique et des Hautes Energies,
CNRS UMR 7589, Universit\'e Pierre et Marie Curie,
Sorbonne Universit\'es, 4 Place Jussieu, 75252 Paris Cedex 05}

\begin{abstract}
A three-terminal Josephson junction consists of three superconductors
coupled coherently to a small nonsuperconducting island, such as a
diffusive metal, a single or double quantum dot. A specific resonant
single quantum dot three-terminal Josephson junction $(S_a,S_b,S_c)$
biased with voltages $(V,-V,0)$ is considered, but the conclusions
hold more generally for resonant semi-conducting quantum wire
set-ups. A simple physical picture of the steady state is developed,
using Floquet theory. It is shown that the equilibrium Andreev bound
states (for $V=0$) evolve into nonequilibrium
Floquet-Wannier-Stark-Andreev (FWS-Andreev) ladders of resonances (for
$V\ne 0$). These resonances acquire a finite width due to multiple
Andreev reflection (MAR) processes. We also consider the effect of an
extrinsic line-width broadening on the quantum dot, introduced through
a Dynes phenomenological parameter. The DC-quartet current manifests a
cross-over between the extrinsic relaxation dominated regime at low
voltage to an intrinsic relaxation due to MAR processes at higher
voltage.  Finally, we study the coupling between the two FWS-Andreev
ladders due to Landau-Zener-St\"uckelberg transitions, and its effect
on the cross-over in the relaxation mechanism. Three important
low-energy scales are identified, and a perspective is to relate those
low-energy scales to a recent noise cross-correlation experiment
[Y. Cohen {\it et al.}, arXiv:1606.08436].
\end{abstract}
\maketitle



\begin{figure}[htb]
\includegraphics[width=.45\textwidth]{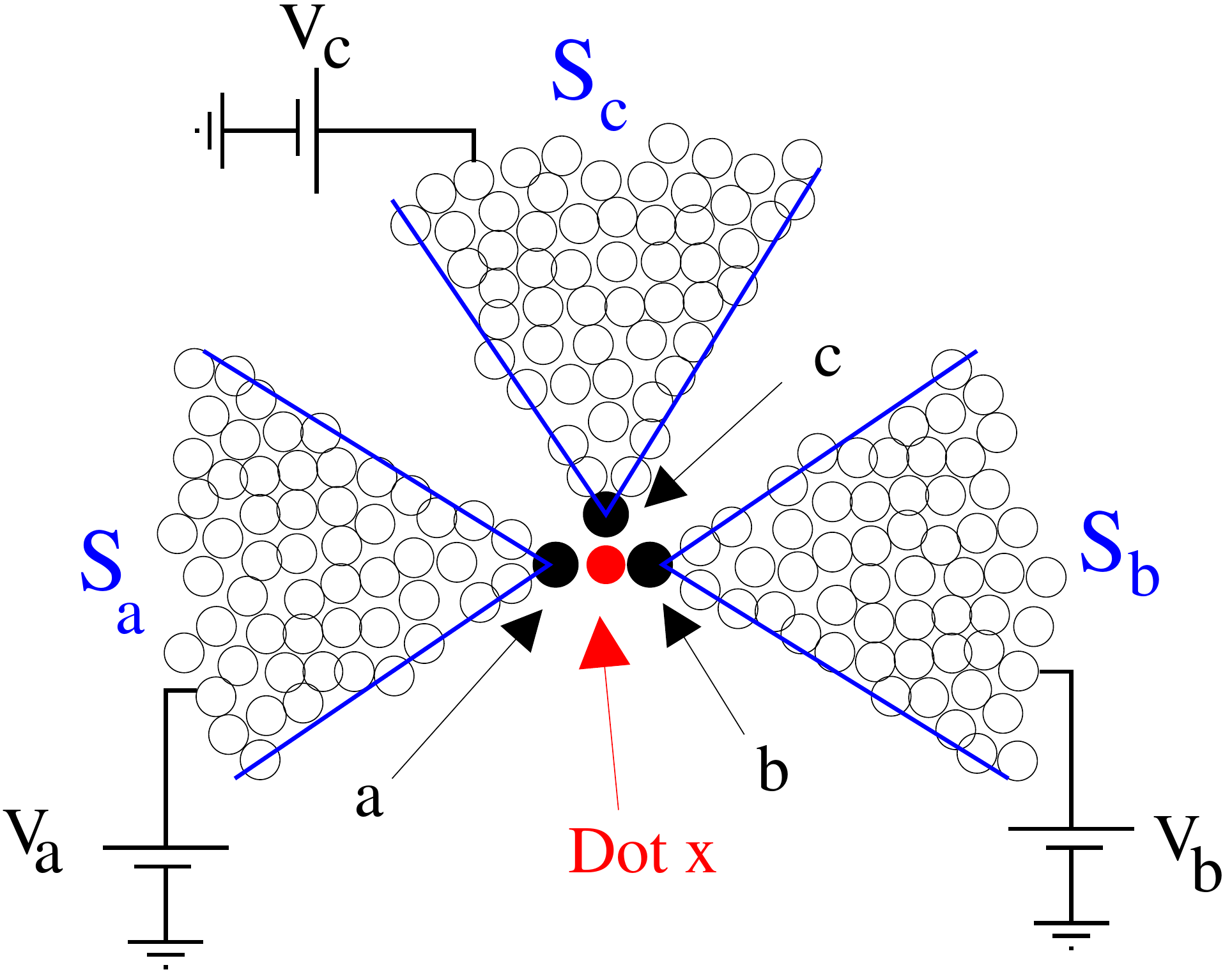}
\caption{{\it The three-terminal set-up:} The figure shows a
  zero-dimensional quantum dot connected to three superconductors. The
  three tunnel couplings between the dot $x$ and $S_a$, $S_b$ and
  $S_c$ take identical values. The voltages are such that $V_a = -V_b
  \equiv V$ (with $V_c = 0$ the reference voltage), and the phases are
  denoted by ($\varphi_a$, $\varphi_b$, $\varphi_c$). The quartet
  phase $\varphi_Q=\varphi_a+\varphi_b-2\varphi_c$ is the relevant
  static phase combination for DC-transport and noise. The quantum dot
  $x$ is shown as a zero-dimensional object because this is the limit
  in which the calculations are carried out. The corresponding
  Hamiltonian is given in Appendix.
\label{fig:1}
}
\end{figure}

\section{Introduction}

A BCS superconductor is characterized by a macroscopic classical
ground state (the condensate) separated by a finite energy gap
$\Delta$ from the first excited states (the superconducting
quasiparticles). Superconducting materials such as those used in
quantum nano-electronics support a tiny density of midgap states
because of electron-electron or electron-phonon processes. The first
investigations of quasiparticle relaxation times were carried out in
the seventies \cite{REF1,Dynes}. Inelastic electron-electron or
electron-phonon interactions in the superconductors produce relaxation
times $\tau_{e-e}$ and $\tau_{e-ph}$ respectively, thus with
corresponding characteristic energies $\hbar/\tau_{e-e}$ and
$\hbar/\tau_{e-ph}$. Those energies for relaxation are accounted for
phenomenologically by the so-called Dynes parameter, in the form of
small imaginary part $\eta_S$ added to the otherwise real-valued
energy of superconducting quasiparticles. The Dynes parameter
corresponds to a single parameter $\eta_S$ for relaxation, accounting
phenomenologically for both electron-electron and electron-phonon
processes. If $\tau_{e-e}$ and $\tau_{e-ph}$ have different orders of
magnitude, then $\eta_S$ can be approximated by the dominant
mechanism, corresponding to taking the maximum between
$\hbar/\tau_{e-e}$ and $\hbar/\tau_{e-ph}$. Once processed with a
small imaginary part $\eta_S$ in energy, BCS theory produces the
expected finite life-time $\hbar/\eta_S$ for quasiparticles. The gap
singularity threshold is also rounded, and, as mentioned above, a
nonvanishingly small density of states is induced inside the
superconducting gap.

A few experimentally relevant situations reveal spectacularly the
importance of taking relaxation into account. Let us mention the noise
of a two-terminal superconducting point contact at equilibrium
\cite{AVERIN,REF3,BOUCHIAT} where the direction of the current flow is
reversed at random by thermal fluctuations, producing telegraph noise
in the Cooper pair current. The noise is inverse proportional to the
Dynes parameter in some window of parameters (the smaller
electron-electron or electron-phonon inelastic scattering, the larger
the signal.)  The following paper brings physical answers in a
situation resembling that of Refs.~\onlinecite{AVERIN}
and~\onlinecite{REF3}, in which electron-electron or electron-phonon
inelastic scattering cannot be discarded. One of the goals of the
present work is to pave the way to a future theoretical investigation
of the recent noise cross-correlation experiment of the Weizmann group
\cite{REF4} (see the concluding Sec.~\ref{sec:planned-study}). In view
of Refs.~\onlinecite{AVERIN,REF3}, this requires handling finite
values for the Dynes parameters as it is done here. High sensitivity
measurements \cite{Pekola1} of Dynes parameter smaller than $\sim
10^{-7}\Delta$ in Aluminum were reported recently, in connection with
the realization of an electron turnstile
\cite{Pekola2,Zanten,Basko2}. Quasiparticle poisoning
\cite{poisoning1,poisoning2,quantronics1,quantronics2} is not captured
explicitly by the present-time version of our model, which is
nevertheless predictive as far as relaxation due multiple Andreev
reflections (MARs) coupling the quantum dot to the semi-infinite
quasiparticle continua is concerned.

Now that the sources of relaxation have been introduced, it is a
matter of fact that an additional terminal does not produce only 
small changes in physics: Novel phenomena take place with three
terminals, which were not there with two terminals. Let us mention the
appearance of a static phase variable for specific voltage
configurations \cite{REF9}, leading to emergence of a Josephson-like
current. Correlations among Cooper pairs is the mechanism for this
DC-superflow in a voltage-biased three-terminal Josephson
junction. Fig.~\ref{fig:1} shows such a set-up, in which three
superconducting electrodes are connected to a single quantum dot.
(Going from three to four terminals is a non trivial step which
produces also unexpected effects: for instance, emergence of
topological effects in a four-terminal Josephson junction
\cite{Nazarov}.) Three-terminal Josephson junctions have been the
subject of experimental investigations in two different groups: with
metallic structures\cite{REF7}, and with semi-conducting quantum wires
\cite{REF4}. Besides metallic structures \cite{Beckmann,Russo,Cadden},
experimental set-ups in the close-by field of Cooper pair splitting
are realized with double quantum dots formed in carbon nanotubes
\cite{Takis} or with semi-conducting quantum wires
\cite{Schonenberger,Heiblum}.
 
As mentioned above, the noise cross-correlations of a voltage-biased
three-terminal Josephson junction were measured recently by the
Weizmann group \cite{REF4}. Current cross-correlations were measured
in this experiment, in addition to providing evidence for the
predicted resonant thresholds related to the gap edge
singularities\cite{REF5,REF6}. Experimental evidence was obtained for
supercurrent of ``quartets'' if the condition $V_a+V_b=0$ on bias
voltages is fulfilled ($V_c = 0$ for the grounded $S_c$, and $V_a =
-V_b \equiv V$ for $S_a$ and $S_b$). The previous pioneer Grenoble
experiment \cite{REF7} had already reported a similar anomaly for the
DC-current response in a metallic structure, but without measurement
of current cross-correlations as it was done in
Ref.~\onlinecite{REF4}. A relevant modeling for the Weizmann group
experiment \cite{REF4} is that of a resonant double quantum dot
connected to three BCS superconductors $S_a$, $S_b$ and $S_c$, and
biased at voltages $V_a$, $V_b$ and $V_c$.

The initial theoretical proposals for Cooper pair splitting
\cite{Byers,Choi,Deutscher,Falci,Buttiker,Martin,Melin-Feinberg} were
based on observation that pairs of entangled electrons can be
extracted from a classical BCS condensate, and be split as two spin-
and energy-entangled quasiparticles transmitted in two different
normal or ferromagnetic leads. Forming virtual states with four
(instead of two) entangled quasiparticles is surprisingly possible in
a three-terminal Josephson junction. A quartet
\cite{REF8,REF9,REF10,REF11,REF12,REF13} consists of four fermions
emanating from two pairs in the grounded $S_c$, which interchange
partners, recombine as two outgoing pairs in $S_a$ and $S_b$, and
eventually disappear in the condensates of $S_a$ and $S_b$. The
overall process is DC, and compatible with energy conservation because
of the specific condition $V_a + V_b = 0$ on the voltage
configuration. However, the supercurrent due to the quartet or
multipair mechanism \cite{REF9} and the corresponding noise
cross-correlations \cite{Quartet-noise} are phase-sensitive, and
individual quasiparticles made of four fermions can hardly be
dissociated from the collective superflow in which those are
embedded. The situation for the quartets has thus its own
specificities in comparison with collective excitations such as
quasiparticles in the quantum Hall effect, having fractional charge.

Voltage-biased three-terminal Josephson junctions are representative
of a class of time-periodic quantum Hamiltonians, which is a matter of
Floquet wave-functions. While realizing this work, numerical data
accumulated, and a stage was reached at which physical pictures and
interpretations emerged from those ``numerical experiments''. In this
respect, the FWS-Andreev viewpoint (Floquet theory combined with band
theory in the context of superconducting set-ups) appears to be the right
starting point. More precisely, the FWS-Andreev resonances are
organized within two alternating FWS-Andreev ladders arising from the
two equilibrium Andreev bound states at positive and negative
energies. Those FWS-Andreev resonances (being nonequilibrium Andreev
resonances) form ``ladders'' because those are replicated to infinity
by Floquet theory, due to the periodic time-dependence of the
Hamiltonian. An expression for the energies of the FWS ladders of
Andreev resonances isolated from the quasiparticle semi-infinite
continua is obtained within a few lines of calculations from
Bohr-Sommerfeld quantization (see Sec.~\ref{sec:II} in the paper, and
Sec.~VI in Supplemental Material). The Floquet energies can be also
evaluated with Green's functions (see Sec.~VII in Supplemental
Material): after lengthy calculations, the fully nonperturbative
Bohr-Sommerfeld quantization result is recovered only in a given
limiting case. This provides evidence for the power of the FWS-Andreev
viewpoint for making analytical calculations, as opposed to Green's
function based on resummed perturbation theory in the coupling between
the dot and the superconducting leads. The FWS-Andreev viewpoint
proceeds with a different perturbation theory in the strength of the
exponentially weak processes coupling the quantum dot to the
quasiparticle continua above the gaps, and in the strength of
Landau-Zener-St\"uckelberg transitions \cite{LZS}. Similarities
between an equilibrium DC-supercurrent (in a phase-biased two-terminal
Josephson junction) and the nonequilibrium DC-current due to the
quartet or multipair mechanism in a three-terminal Josephson junction
are considered in the following. This discussion is based on the
discovery reported below of several relevant low-energy scales in the
current, and in the FWS spectrum of Andreev resonances.

The paper is organized as follows. The FWS-Andreev viewpoint is
presented in Sec.~\ref{sec:II}. Numerical calculations are presented
in Sec.~\ref{sec:III}. Summary and perspectives are provided in the
concluding Sec.~\ref{sec:V}. The Hamiltonian is provided in
Appendix.

More technical discussions are relegated to Supplemental Material:
details on the demonstration of the adiabatic theorem (Secs.~I and~II
in Supplemental Material), principle of the code with finite
$\eta_{dot}/\Delta$ (Sec.~III in Supplemental Material). The failure
of the Keldysh dressing algorithm with fully discrete spectrum is
demonstrated (Sec.~IV in Supplemental Material). The principle of
another code based on the Floquet-Lippmann-Schwinger dressing
algorithm is given in Sec.~V of Supplemental Material. Sec.~VI in
Supplemental Material presents a complementary and more general
introduction to FWS ladders, not necessarily in the context of
superconductivity. The Rabi frequencies are calculated in Sec.~VII of
Supplemental Material from the rotating wave approximation, and from
microscopic Green's functions. A detailed physical discussion of the
nonstandard generalized Dynes parameter is presented in Sec.~VIII of
Supplemental Material.

\section{Floquet-Wannier-Stark-Andreev viewpoint}
\label{sec:II}

\subsection{Two-terminal junctions}
\label{sec:IIA}
\begin{figure}[htb]
\includegraphics[width=.45\textwidth]{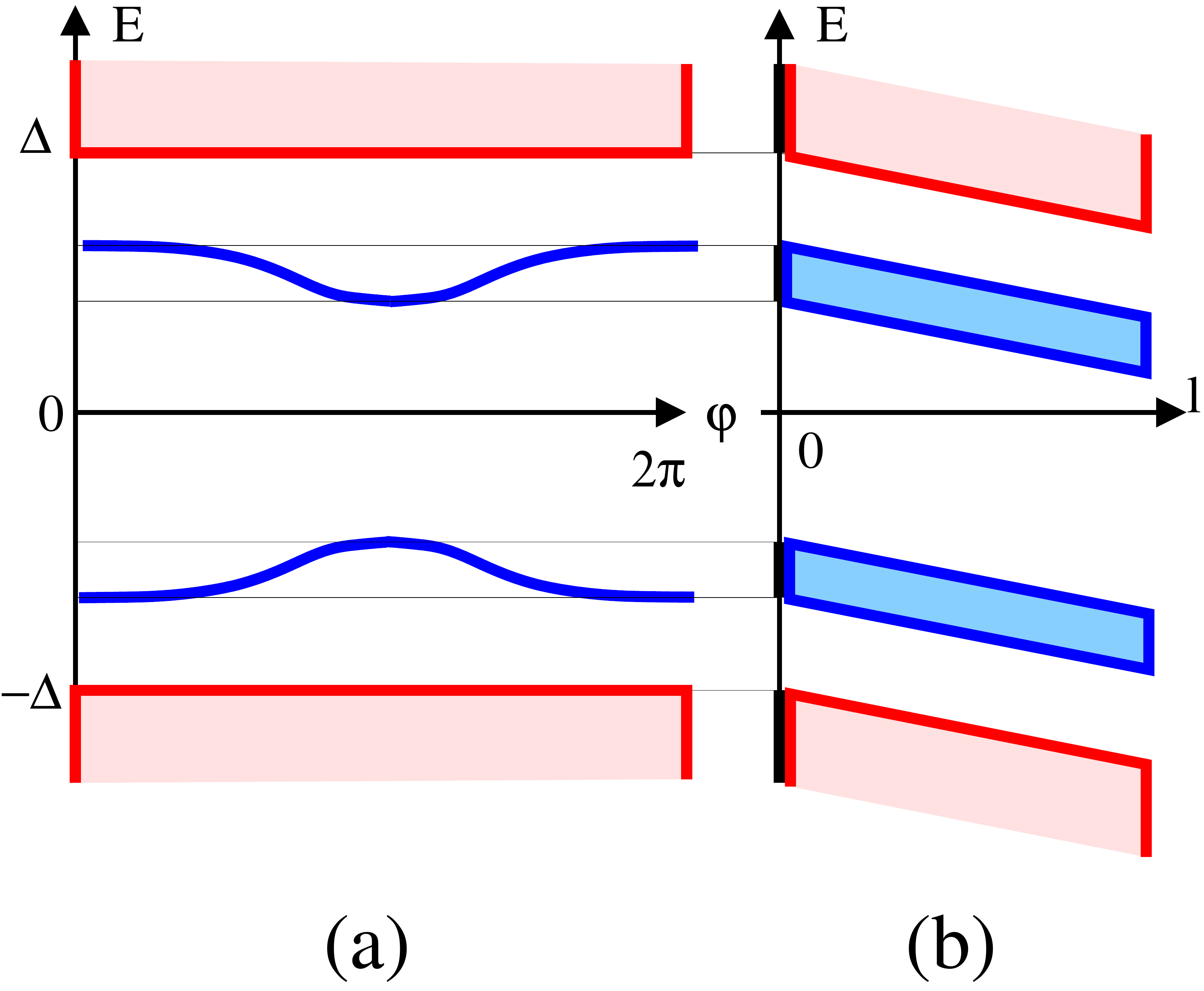}
\caption{{\it The FWS-Andreev tilted band picture:} Panel a shows the
  schematics for the energy-phase relation of a two-terminal Josephson
  junction with an embedded quantum dot supporting a single energy
  level, with two Andreev levels at opposite energies. Panel b shows
  how the tilted band picture in the $(l, E)$ plane is related to the
  energy-phase relation, where $l=(N_a-N_b)/2$ (with $N_a$ and $N_b$
  the number of fermions transmitted in $S_a$ and $S_b$). The
  quasiparticle continua are shown in red. The dispersion of the two
  Andreev bound states is shown in blue on panel a. The filled blue
  region on panel b shows the FWS-Andreev bands. The noncolored
  regions are classically forbidden. It is supposed on this figure
  that the two superconducting leads are asymmetrically coupled to the
  quantum dot, therefore providing a gap between Andreev bound states
  for all values of the phase difference.
\label{fig:2}
}
\end{figure}

Floquet theory can be introduced for a general time-periodic
Hamiltonian \cite{Shirley65,Sambe73} (see also Sec.~VI in Supplemental
Material), but its physical meaning becomes rather transparent in the
context of superconducting junctions. The Josephson relation implies a
time-periodic Hamiltonian (see Appendix) for the Hamiltonian of the
three-terminal set-up in Fig.~\ref{fig:1}). The gauge can be chosen in
such a way as to get rid of the time dependence of the Hamiltonian,
and to handle instead (now for a two-terminal junction):
\begin{equation}
  \label{eq:H2TER}
  \tilde{\cal H}=\hat{\cal H}_a
  +\hat{\cal H}_b+ \hat{\cal H}_{a-b}
  -eV\left(\hat{N}_a-\hat{N}_b\right)
  ,
\end{equation}
where $\hat{\cal H}_a$ and $\hat{\cal H}_b$ are the BCS Hamiltonians
of the two superconducting leads $S_a$ and $S_b$ coupled by the tunnel
term $\hat{\cal H}_{a-b}$. The term $-eV(\hat{N}_a-\hat{N}_b)$ is the
energy of the $N_a-N_b$ fermions which have been transmitted between
the two electrodes. Eq.~(\ref{eq:H2TER}) is supplemented with the
commutation relation
\begin{equation}
  \label{eq:commutation}
\left[\hat{N}_a - \hat{N}_b , \frac{\hat{\varphi}_a -
    \hat{\varphi}_b}{2} \right]=i
,
\end{equation}
which makes this problem quantum mechanical. It is illustrative to
plot $l=(N_a-N_b)/2$ on the $x$-axis and energy on the $y$-axis (see
Fig.~\ref{fig:2}), not taking Eq.~(\ref{eq:commutation}) into account
in a first step. Two semi-classical energy bands are obtained for
$l=0$ (spanned by the two equilibrium Andreev bound states at positive
and negative energies), and increasing $l=(N_a-N_b)/2$ produces the
tilt shown in Fig.~\ref{fig:2}, due to the term $2eVl=eV(N_a-N_b)$
subtracted to the Hamiltonian. The semi-infinite quasiparticle
continua are also shown in this tilted band picture.

Now, quantum mechanics is introduced. The Floquet wave-functions are
evaluated and the corresponding resonant levels are quantized. The
FWS-Andreev wave-functions are located essentially in the classically
allowed regions (filled blue regions in Fig.~\ref{fig:2}b), and those
are evanescent in the classically forbidden regions. Considering first
two isolated FWS-Andreev bands, the expression of the resonant level
energies is obtained from Bohr-Sommerfeld semi-classical quantization
on the Floquet wave-function. Tunneling into quasi-particle continua will produce
a finite line-width for the FWS-Andreev resonances.

Eigenstates of the extended Hamiltonian $\pm E(\hat{\varphi})-2eV
\hat{l}$ are obtained, where $\hat{\varphi} = \hat{\varphi}_a -
\hat{\varphi}_b$. In this expression, $E(\varphi)$ is the energy-phase
relation of the upper bound state, and $\pm$ refers to Andreev bound
states at positive and negative energies respectively. The term
$-2eV\hat{l}$ is the energy of the number
$\hat{l}=(\hat{N}_a-\hat{N}_b)/2$ of pairs which have been transmitted
during the quantum process. The trick is to note that $\hat{l}$ and
$\hat{\varphi}$ are canonically conjugate variables; see
Eq.~(\ref{eq:commutation}), which is equivalent to
$\left[\hat{l},\hat{\varphi}\right]=i$. The operator $\hat{l}$ is then
exactly identical to $\hat{l}=i\partial/\partial \varphi$. The
Schr\"odinger equation in extended space takes the form of a
first-order differential equation for the variable $\varphi$ (see
Sec.~VI in Supplemental Material). The solution for the wave-function
takes the form
\begin{equation}
  \psi_\pm(\varphi)\sim\exp\left(\frac{i}{2eV}
  \int_0^\varphi(E\mp E(\varphi'))d\varphi'
\right)
.
\end{equation}
Imposing $2\pi$-periodicity leads to the following quantized energies
of the FWS-Andreev resonant levels:
\begin{eqnarray}
\label{eq:1}
E_j &=& 2 eV j+ \langle E \rangle\\
\label{eq:2}
E'_{j'} &=& 2 eV j' - \langle  E \rangle
,
\end{eqnarray}
where $j$ and $j'$ are integers, and $\langle E \rangle$ denotes the
average over the phase of the energy of the upper Andreev bound state:
\begin{equation}
  \langle E \rangle=\frac{1}{2\pi}\int_0^{2\pi} E(\varphi) d\varphi
  .
\end{equation}
The minus sign in Eq.~(\ref{eq:2}) is due to the opposite energies of
the Andreev bound states at negative and positive energies; their
energies $\pm\langle E \rangle$ averaged over the phase are also
exactly opposite. This argument and additional comments are presented
in Sec.~VI of Supplemental Material in the case of a single
band. Similar equations were obtained long ago in classical papers on
Wannier-Stark theory (see for instance Ref.~\onlinecite{REF14}).

It is important to note that for two weakly coupled Wannier-Stark
ladders, the average energy $\langle E \rangle$ depends on $V$. This
corresponds to dressing each eigenstate in a given ladder by virtual
transitions to the other ladder. These processes can be captured in a
perturbative expansion in $V$. This is to be contrasted with
Landau-Zener-St\"{u}ckelberg transitions, which appear as tunneling
processes between ladders, therefore giving contributions which are
analytic functions of $1/V$ (see below).  In Eqs.~(\ref{eq:1}) and
(\ref{eq:2}), only the first processes are taken into account, through
the analytic dependence of $\langle E \rangle$ as a function of $V$.

\begin{figure}[htb]
  \includegraphics[width=.38\textwidth]{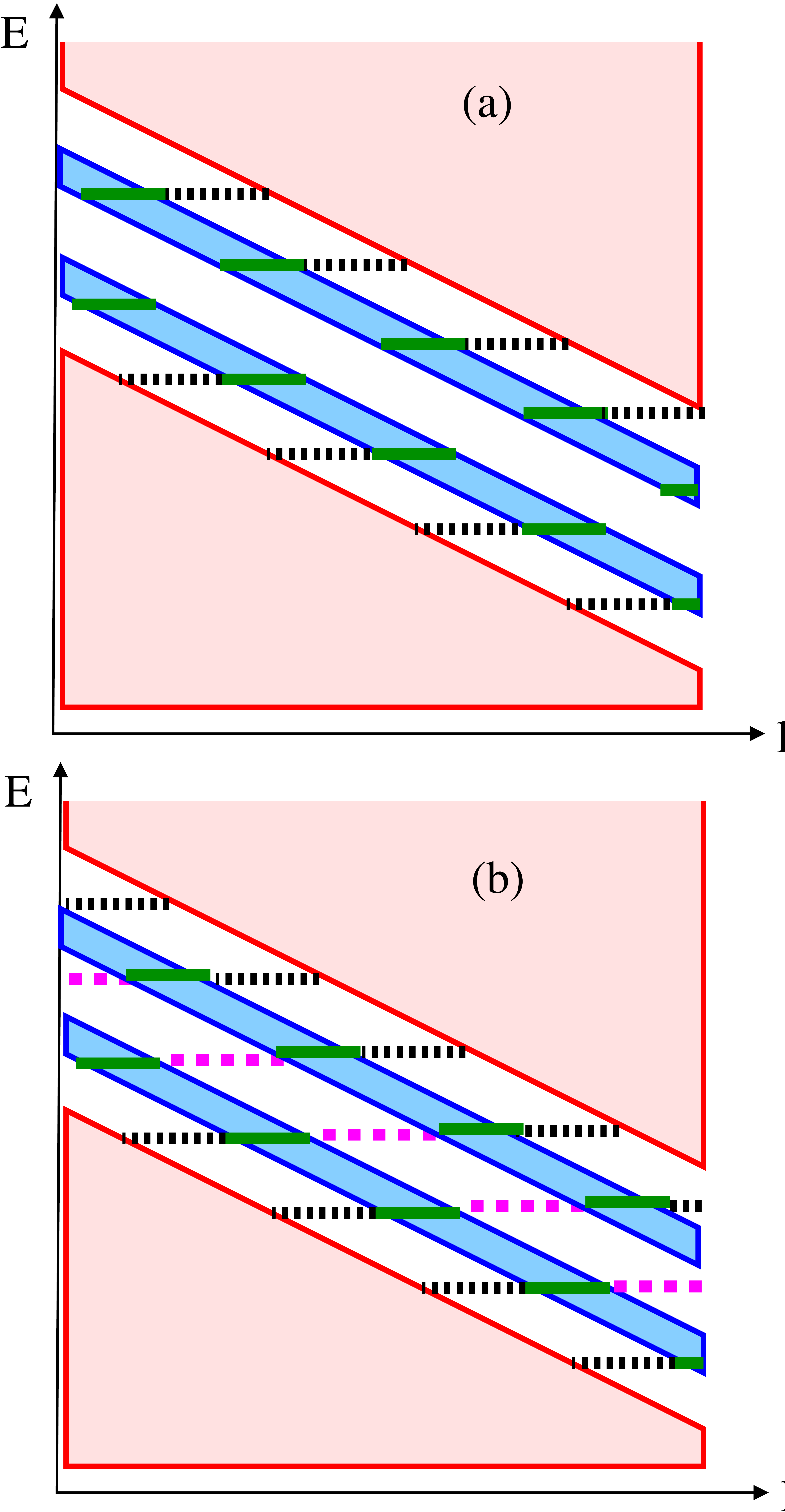}
\caption{{\it FWS-Andreev ladders:} The figure shows schematically the
  FWS-Andreev ladders for a two-terminal Josephson junction in the
  $(l, E)$ plane, where the auxiliary variable $l=(N_a-N_b)/2$ is
  conjugate to the phase difference $\varphi_a-\varphi_b$. Panel a
  shows FWS-Andreev ladders with misaligned FWS-Andreev resonances
  (green solid lines). Tunneling is then essentially with the
  semi-infinite quasiparticle continua (dashed back lines). Panel b
  shows schematically aligned FWS-Andreev resonances, with also
  interband tunneling in this case. The energies of the FWS-Andreev
  resonances are given by Eqs.~(\ref{eq:1}) and (\ref{eq:2}). The
  length of the solid lines shows the extent of the wave-functions
  along the axis of the auxiliary variable $l$. The wave-functions can
  be delocalized on both ladders in the case of aligned FWS-Andreev
  resonances, because of interband tunneling through the classically
  forbidden region (panel b). Interband tunneling produces energy
  level repulsion between hybridized FWS-Andreev resonances in this
  case.
\label{fig:3}
}
\end{figure}

The FWS-Andreev resonance energies defined by Eqs.~(\ref{eq:1})
and~(\ref{eq:2}) are shown by solid horizontal lines on top of the
band structure of the extended Hamiltonian (see
Fig.~\ref{fig:3}). Fig~\ref{fig:3}a shows two ladders with FWS-Andreev
resonances being significantly ``misaligned'' in energy (tunneling is
then essentially with the quasiparticle continua). Fig.~\ref{fig:3}b
corresponds to FWS-Andreev resonances almost ``aligned'' in energy
(with possibility of tunneling between the two Andreev resonances
belonging to different ladders). In all cases, tunneling proceeds
through a classically forbidden region, which has a length [in the
  auxiliary variable $l=(N_a-N_b)/2$] inverse proportional to the tilt
(the tilt is proportional to bias voltage). The theory of tunnel
effect states that tunneling decreases exponentially with the length
of the barrier. Tunneling is thus exponentially small in the inverse
of bias voltage. Tunneling between the FWS-Andreev resonances and the
continua is consistent with the voltage-dependence of the FWS-Andreev
line-width broadening, which will be evaluated numerically in
Sec.~\ref{sec:III}.

\subsection{Three-terminal junctions}
\begin{figure}[htb]
\includegraphics[width=.3\textwidth]{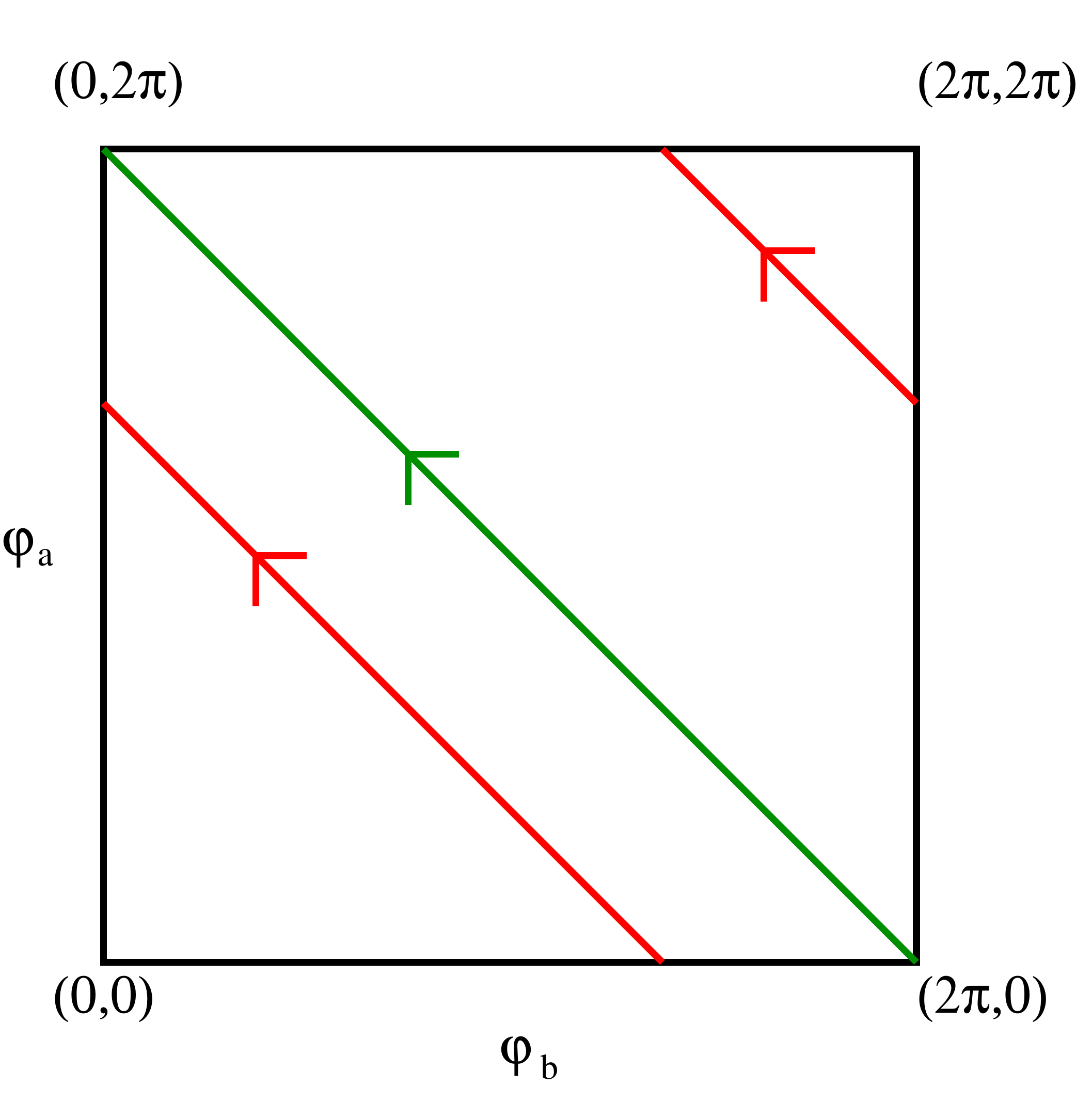}
\caption{The figure shows, for a three-terminal Josephson junction,
  the trajectories of the phases ($\varphi_a(t)$, $\varphi_b(t)$) with
  a vanishingly small quartet phase $\varphi_Q=0$ (green), and with a
  generic quartet phase $\varphi_Q\ne 0$ (red line). The
  time-dependence of the phases are given by
  $\varphi_{a,b}(t)=\varphi_{a,b}(0)\pm 2eVt/\hbar$, where opposite
  voltages $V_{a,b}=\pm V$ are applied on $S_{a,b}$ with respect to
  the grounded $S_c$.
    \label{fig:intro-3ter}
}
\end{figure}

Now, remarks are provided on a generalization to three superconducting
terminals connected to a single dot (see Fig.~\ref{fig:1}). In this
case, the equilibrium Andreev bound states depend on the two phase
variables $\varphi_a$ and $\varphi_b$. With opposite bias voltages
$\pm V$ on $S_a$ and $S_b$ ($S_c$ being grounded), the time evolution
of those phases is such that $\varphi_a(t)=\varphi_a(0)+2eVt/\hbar$
and $\varphi_b(t)=\varphi_b(0)-2eVt/\hbar$ (and $\varphi_c=0$ is the
reference phase). As mentioned in the Introduction, the quartet phase
$\varphi_Q=\varphi_a(t)+\varphi_b(t)=\varphi_a(0)+\varphi_b(0)$ is
static if $V_a+V_b=0$, and the phase difference
$\varphi_a(t)-\varphi_b(t)$ is winding in time:
$\varphi_a(t)-\varphi_b(t)=4eVt/\hbar$. The trajectories of the phases
in the $(\varphi_a,\varphi_b$) plane are shown in
Fig.~\ref{fig:intro-3ter} for $\varphi_Q=0$ (green line) and for a
generic $\varphi_Q\ne 0$ (red line). For the former, the gap between
Andreev bound states closes at two opposite values of the phases,
located in between $(0,0)$ and $(\pi,\pi)$ and, for the opposite
phases, in between $(0,0)$ and $(-\pi,-\pi)$ (see
Ref.~\onlinecite{Akhmerov}). On the contrary, the gap between Andreev
bound states does not close in the generic case $\varphi_Q\ne 0$
considered in the forthcoming Sec.~\ref{sec:III} (red line in
Fig.~\ref{fig:intro-3ter}). The auxiliary variable $l=(N_a-N_b)/2$ is
conjugate to the fast variable $\varphi_a-\varphi_b$. The FWS-Andreev
viewpoint is identical to Fig.~\ref{fig:3}, but now this figure is
also parameterized by the quartet phase~$\varphi_Q$. Eqs.~(\ref{eq:1})
and~(\ref{eq:2}) still hold for three-terminals, where $\langle E
\rangle$ is the average over $\varphi_a-\varphi_b$ of the energy of
the upper bound state. For a three-terminal set-up, $\langle E
\rangle$ is also parameterized by the quartet phase $\varphi_Q$.

\subsection{Self-induced Rabi resonances}

The first theoretical work \cite{REF8} addressing phase-coherent
three-terminal Josephson junctions coined those quartets and
higher-order resonances as ``self-induced Shapiro steps'' appearing
for rational ratios between bias voltages. Going now one step further,
it is found here that, at fixed commensuration ratio $V_b/V_a=p/q$
(with $p$ and $q$ two integers), voltage-parameterized ``self-induced
Rabi resonances'' are obtained between FWS-Andreev resonances belonging to
different sub-bands. However, self-induced Rabi resonances are not
specific to three superconducting terminals: on the contrary, similar
Rabi resonances are expected also in a set-up consisting of two
superconductors connected to a quantum dot. Three-terminal Josephson
junctions at the quartet resonance offer the quartet phase as an
additional control parameter.

Coming back to a three-terminal Josephson junction with
$V_a=-V_b\equiv V$, it is deduced from Eqs.~(\ref{eq:1})
and~(\ref{eq:2}) that self-induced Rabi resonances appear at
voltages
\begin{equation}
  \label{eq:Omega}
  2eVj+\langle E \rangle = 2eVj'-\langle E \rangle
  ,
\end{equation}
with $j$ and $j'$ two integers. In the low voltage limit, a sequence of Rabi resonances at
voltages
\begin{equation}
  V_k=\frac{V_1}{k}
  \label{eq:Vk}
\end{equation}
is expected, with $k$ an integer, and $eV_1=\lim_{V \rightarrow 0} \langle E \rangle$. In the
limit of small coupling $\Gamma/\Delta\ll 1$ between the dot and the
superconducting leads, this leads to $eV_R\simeq \Gamma$. Tunneling
between the dot and each of the superconducting electrodes is
parameterized in a standard way by $\Gamma=J^2/W$, with $J$ the
corresponding hopping matrix element and $W$ the band-width.

\begin{figure}[htb]
  \includegraphics[width=.45\textwidth]{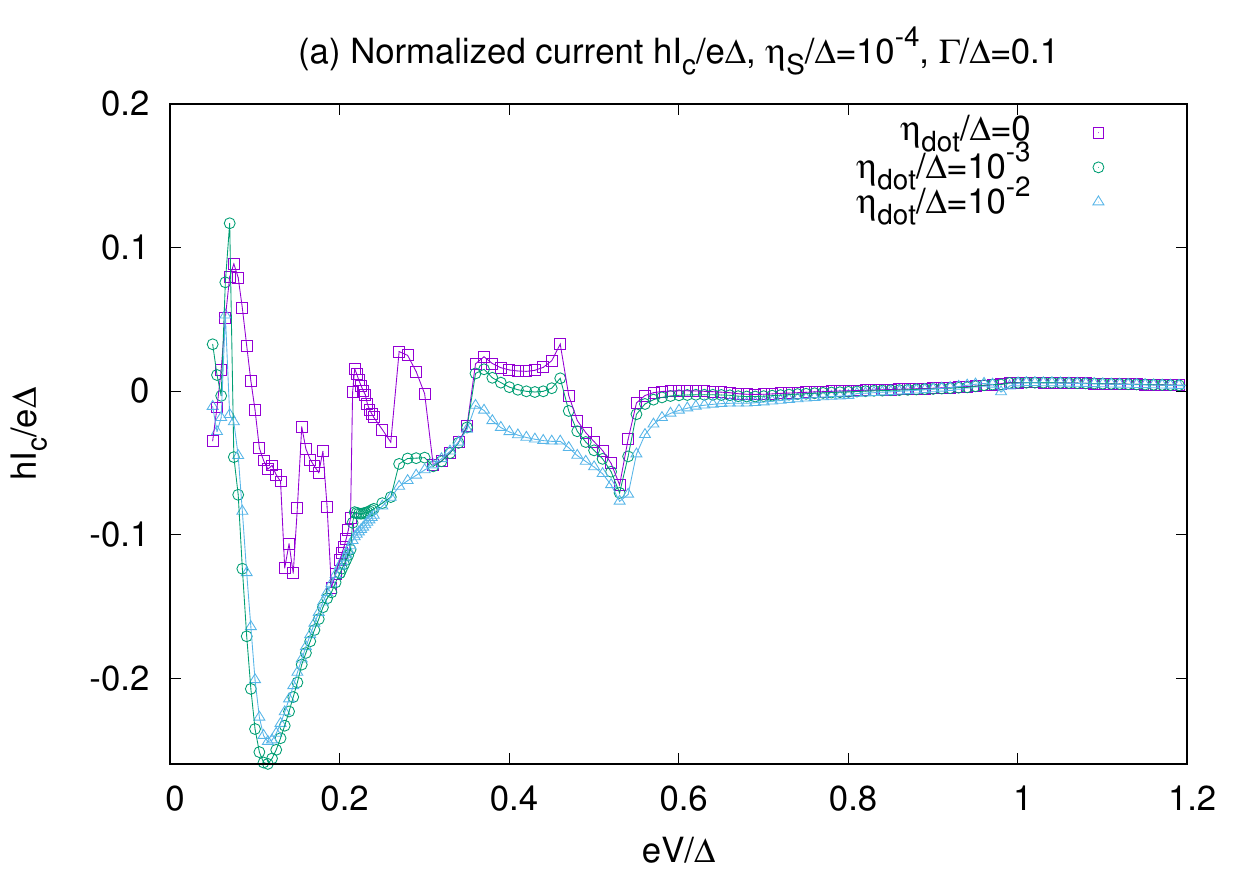}

  \includegraphics[width=.45\textwidth]{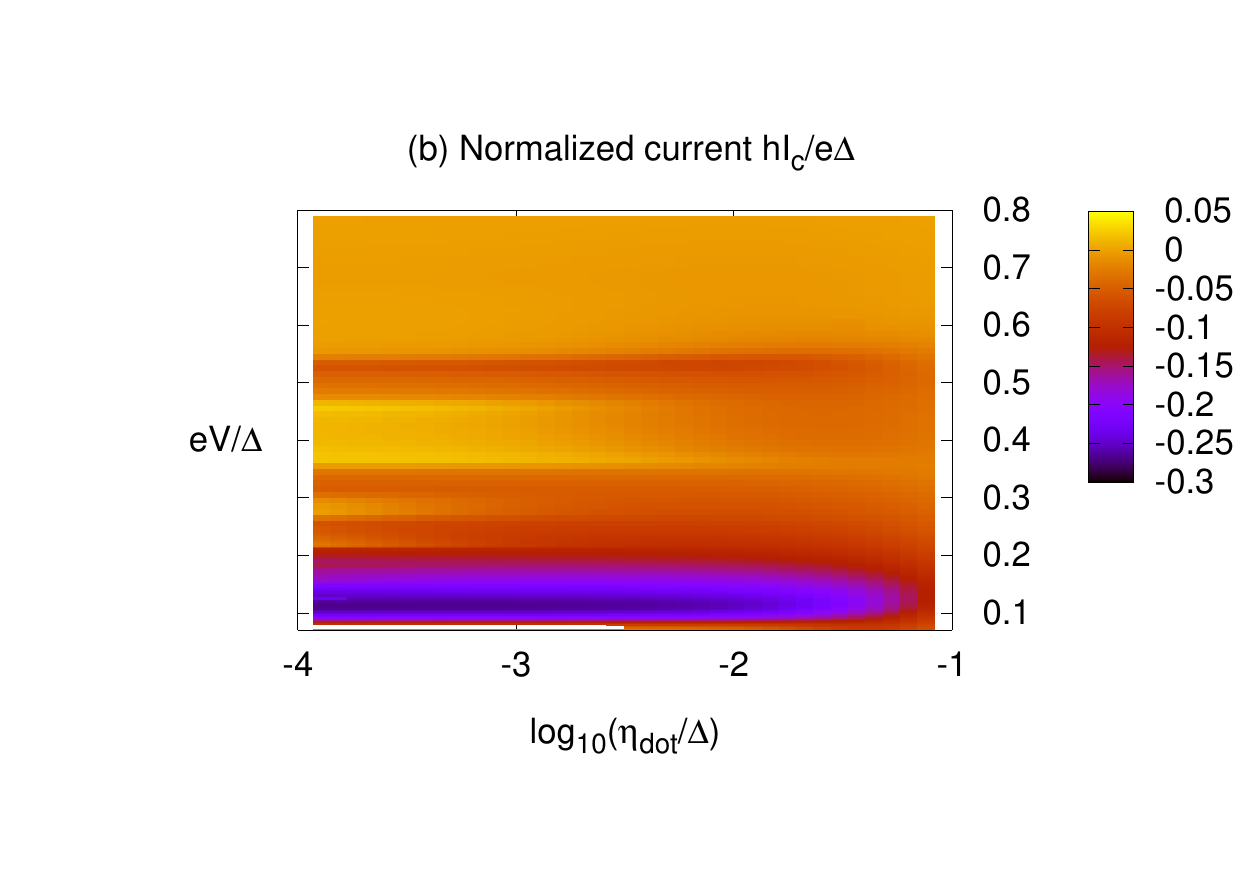}
\caption{{\it Variations of the current $I_c$ through lead $S_c$:}
  Panel a shows the voltage dependence of the normalized current
  $hI_c/e\Delta$ as a function of normalized voltage $eV/\Delta$ for
  the values of $\eta_{dot}/\Delta$ indicated on the figure. Panel b
  shows $h I_c / e \Delta$ in the plane of the variables
  ($\log(\eta_{dot}/\Delta)$, $eV/\Delta$).
\label{fig:4}
}
\end{figure}

\subsection{Generalized Dynes parameter $\eta_{dot}$}
A finite bias voltage makes the FWS-Andreev resonances qualitatively
different from Andreev resonances at equilibrium. For instance, MARs
due to finite voltage contribute to the width of FWS-Andreev
resonances. MARs produce only tiny relaxation at sufficiently low
voltage (exponentially small in $\Delta/eV$), thus indicating that
FWS-Andreev ladders of resonances are robust against relaxation due to
the coupling by MARs to the semi-infinite quasiparticle
continua. However, if one wants to make experiments or to put this
problem on a computer as it is done below, other mechanisms for
relaxation have to be taken into account, which will be dominant once
MARs become too weak at low bias voltage. From the point of view of
numerical calculations, relaxation is encoded in the Dynes parameter
$\eta_S$ in the superconductors. Electron-phonon scattering is a
natural candidate for relaxation on the quantum dot (see also
Sec.~VIII of Supplemental Material). Once the Dynes parameter $\eta_S$
has been introduced in the superconductors, then the generalized Dynes
parameter $\eta_{dot}$ has to be also included on the quantum dot,
because of a question of internal consistency of the description
(phonons operate both in the superconducting leads and on the quantum
dot). It is then obvious that, for a resonant quantum dot,
$\eta_{dot}$ is much more efficient than $\eta_S$ as a relaxation
mechanism (See Sec.~\ref{sec:III} for the numerical results and
Sec.~VIII in Supplemental Material for an explanation based on an
analytical Green's function calculation). Transport theory with finite
$\eta_{dot}$ is more challenging than for $\eta_{dot}=0$, but a
mathematically exact solution for the current can be well obtained at
finite $\eta_{dot}$ (see Secs.~I, II and III in Supplemental
Material). The authors are aware that a finite $\eta_{dot}$ is against
recipes in the literature, and more especially against our previous
work \cite{REF11}: The limit $\eta_{dot}=0^+$ was taken, and, in
practise, $\eta_{dot}=0$ was implemented without questioning why
$\eta_{dot}=0$ is equivalent to $\eta_{dot}=0^+$. The smoking gun that
nontrivial physics is related to a finite $\eta_{dot}\ne 0$ is that
adiabatic theorem breaks down severely for a three-terminal Josephson
junction with two levels at energies $\pm \epsilon_0$ (in this case,
the currents calculated with $\eta_{dot}=0^+$ are not equal to those
for $\eta_{dot}=0$) (See Sec.~IID of Supplemental Material). The
situation is not so drastic for a single quantum dot having a level a
zero energy: adiabatic theorem can well be demonstrated in this case
(see Secs.~I and II of Supplemental Material), which provides evidence
that our previous calculations \cite{REF11} for $\eta_{dot}=0$ are
mathematically correct in the sense that the currents take identical
values for $\eta_{dot}=0$ and $\eta_{dot}=0^+$. However, our previous
work \cite{REF11} leaves open central questions such as those related
to the speed of convergence of the current to its adiabatic limit
value: The latter is found here to be exponentially small in
$\Delta/V$, which, in essence, is due to the emergence of those
low-energy/long-time scales of interest here. Indeed, the
characteristic $\eta_{dot}^*$ for cross-over in the current to the
adiabatic limit (corresponding $\eta_{dot}\ll\eta_{dot}^*$) receives
the interpretation of the inverse of an intrinsic characteristic time
scale.

On a more physical basis, some modes of the quantum dot are almost
isolated from their environment, which explains long relaxation
times. For instance, heat transport is expected to be very slow at low
bias voltage for the set-up in Fig.~\ref{fig:1}, because the
environment of the quantum dot consists of three BCS superconductors
which do not propagate entropy over large distances (see the
discussion in the concluding Sec.~\ref{sec:concluC}). Supplemental
Material offers large amount of technical details intended to unveil
the theoretical framework behind the numerical results presented now.

\section{Numerical calculations}
\label{sec:III}
The following calculations are carried out with a resonant quantum dot
connected to three superconducting leads $S_a$, $S_b$ and $S_c$ biased
at $V_{a,b} = \pm V$ and $V_c = 0$ respectively, in the presence of a
generic finite value for the static quartet phase $\varphi_Q \ne
0$. The starting point is the Hamiltonian given in Appendix. The
(Keldysh) Green's functions are calculated from the Dyson(-Keldysh)
equations in extended space, and the currents are expressed in terms
of the Dyson-Keldysh Green's function. The codes with $\eta_{dot}\ne
0$ are based on implementation of a self-consistency loop on the bare
populations of the quantum dot. All numerical results presented below
are obtained with recursive Green's functions in energy (see
Refs.~\onlinecite{REF16,Cuevas-noise}). Technical details on the
principle of the codes are provided in Supplemental Material (see more
especially Sec.~III in Supplemental Material for a discussion of
$\eta_{dot}\ne 0$).

The variations of current $I_c$ through lead $S_c$ are shown in
Fig.~\ref{fig:4} as a function of normalized voltage $eV/\Delta$ and
normalized generalized Dynes parameter $\eta_{dot}/\Delta$, for
$\Gamma/\Delta = 0.1$. As mentioned above, the contact transparencies
are parameterized by $\Gamma=J^2/W$, with $J$ the hopping matrix
element between the dot and the superconducting leads, and $W$ the
band-width.  It was verified on a few examples that the following
results are compatible with those obtained previously \cite{REF10} for
$\eta_{dot}/\Delta=0$. A technical discussion is summarized in
Secs.~I, II and III in Supplemental Material, on how our calculations
with a tiny $\eta_{dot}/\Delta=0^+$ relate to those with
$\eta_{dot}/\Delta=0$ in our previous work \cite{REF10}. The
variations of $I_c$ with $eV /\Delta$ (see Fig.~\ref{fig:4}a) look
much smoother than those obtained for $\eta_{dot}/\Delta=0$ once a
tiny $\eta_{dot} /\Delta = 10^{-3} \div 10^{-2}$ is included. The
currents are evaluated according to the self-consistent procedure
presented in Sec.~III of Supplemental Material. The first of the
self-induced Rabi resonances [see Eq.~(\ref{eq:Vk}) with $k=1$] is in
agreement with the maximum at $eV/\Delta\simeq 0.1$ in the voltage
dependence of the normalized current $h I_c/e\Delta$ (see
Fig.~\ref{fig:4} for $\eta_{dot}/\Delta=10^{-3}$, $10^{-2}$ and
$\Gamma/\Delta = 0.1$). This Rabi resonance was predicted in
Sec.~\ref{sec:II} at $eV_R = \langle E \rangle \simeq \Gamma$ if
$\Gamma/\Delta \ll 1$. Fig.~\ref{fig:4} does not fully match this
condition of tunnel contacts, because $\Gamma/\Delta=0.1$ is not a
very small number. A resonance appears in Fig.~\ref{fig:4}a at voltage
slightly larger than $V_R / \Delta = 0.1$, and that is also visible in
Fig.~\ref{fig:4}b, featuring the normalized current $hI_c/e\Delta$ (in
color-scale) in the plane of parameters ($\log(\eta_{dot} /\Delta)$,
$eV /\Delta$). To conclude the discussion of those first numerical
observations, ultra-sensitivity on a tiny $\eta_{dot}/\Delta$ (see
Fig.~\ref{fig:4}) is a first numerical evidence for emergence of
small-energy/long-time scales: a small cross-over value $\eta_{dot}^*$
for convergence towards adiabatic limit in
$I_c(\log(\eta_{dot}/\Delta))$ implies a long time
$\hbar/\eta_{dot}^*$. In addition, it is found here that the voltage
scale $V_R$ controls the voltage-dependence of the current once a
small $\eta_{dot}/\Delta$ is included.

\begin{figure}[htb]
\includegraphics[width=.45\textwidth]{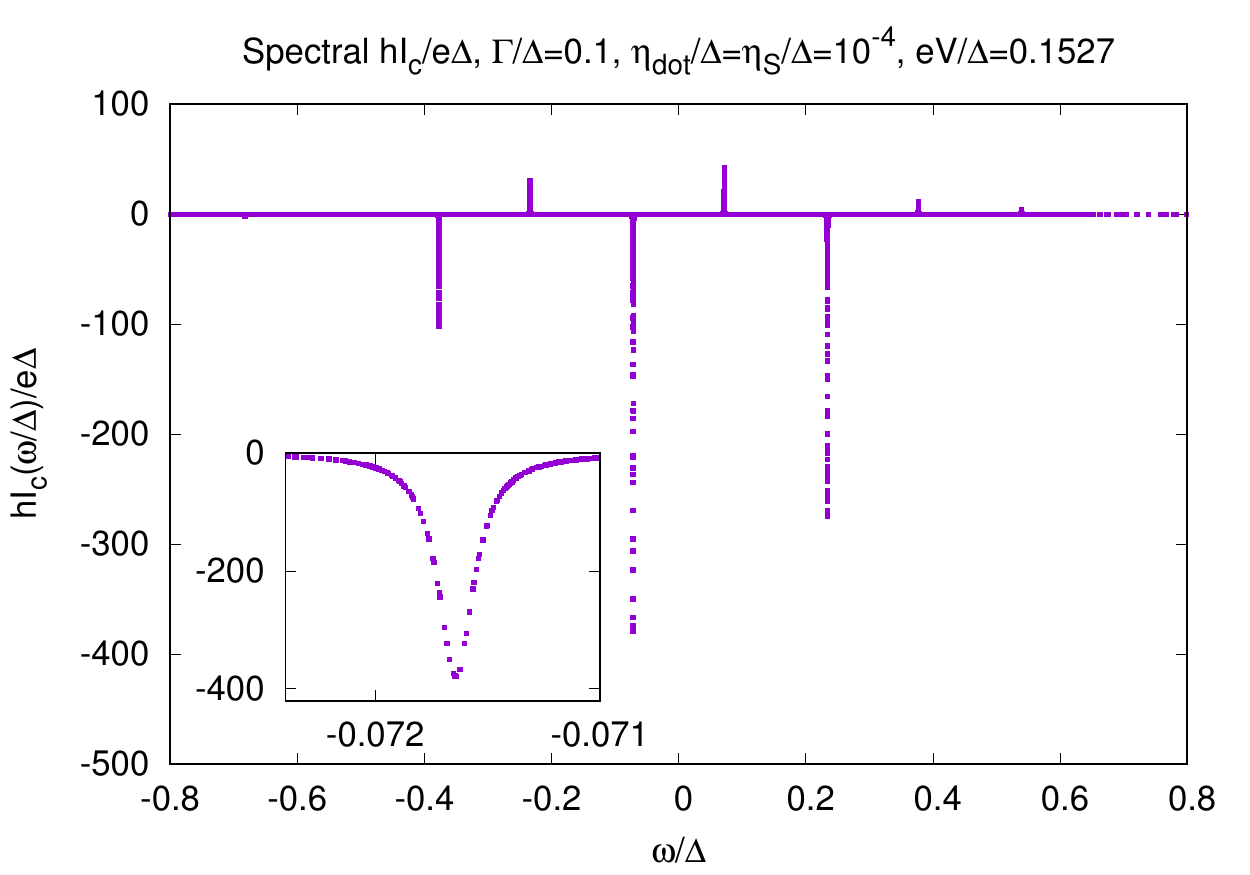}
\caption{{\it Spectral current $I_c$:} The figure shows the
  frequency~$\omega$ dependence of the normalized spectral current $h
  I_c(\omega/\Delta)/e\Delta$. The insert (with the same axis labels
  as the main panel) shows the same quantity in the vicinity of a
  specific resonance.
\label{fig:5}
}
\end{figure}

\begin{figure}[htb]
  \includegraphics[width=.45\textwidth]{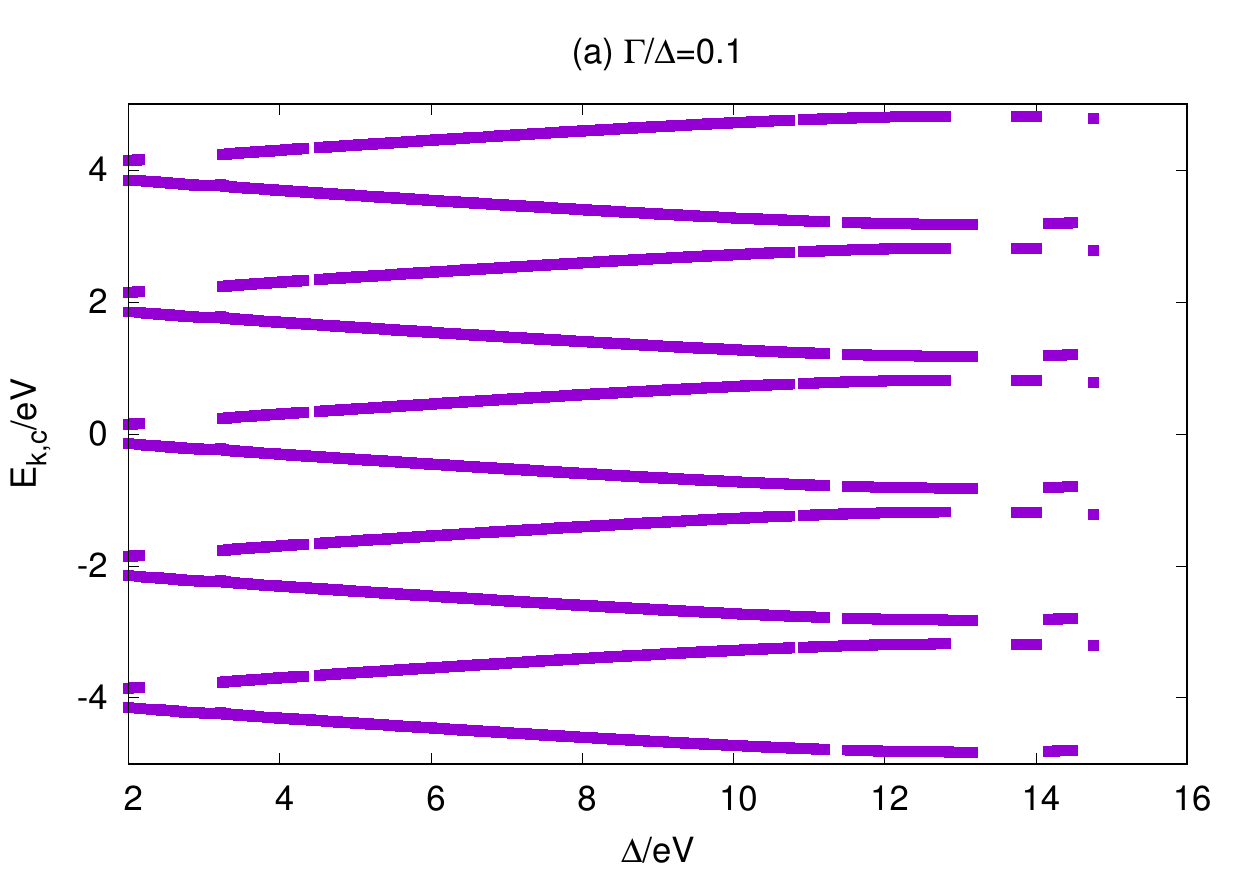}

  \includegraphics[width=.45\textwidth]{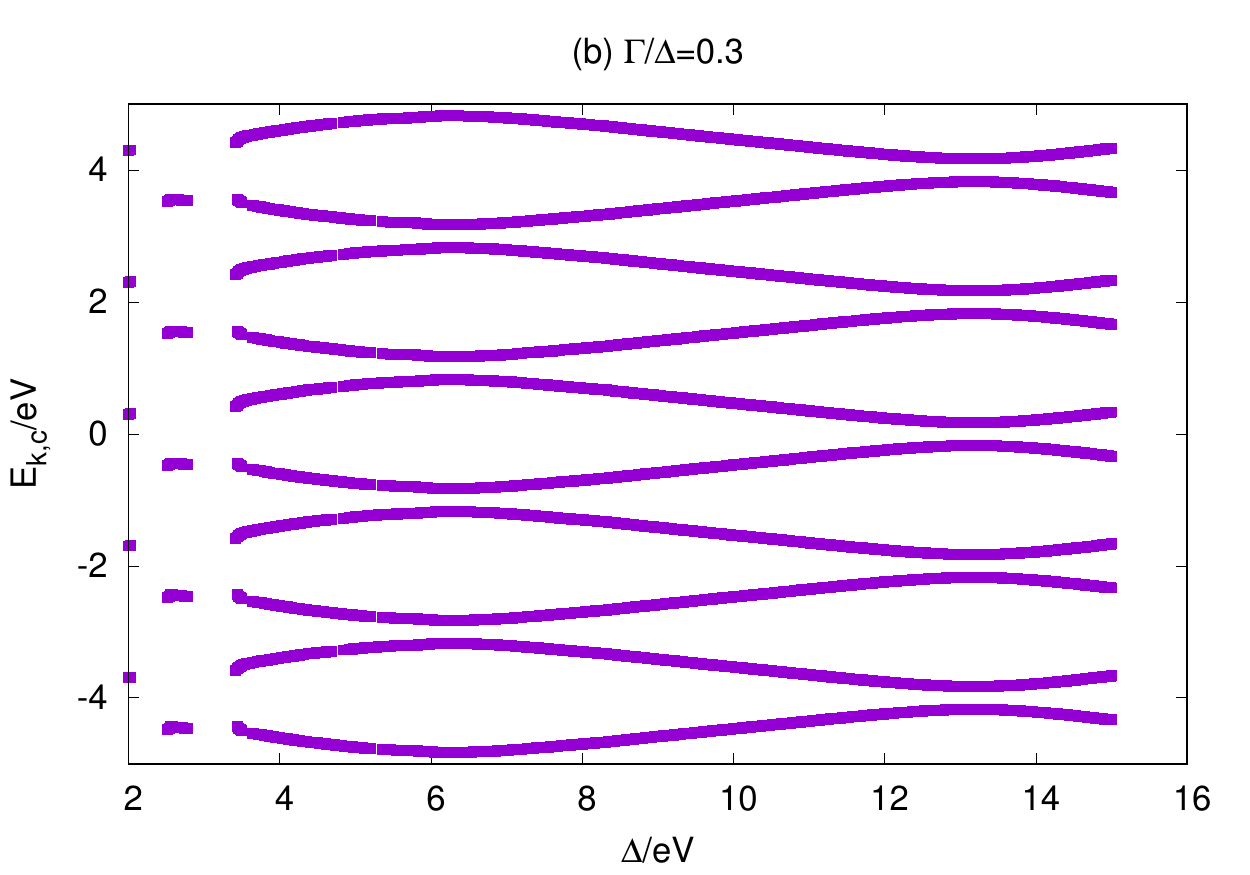}
\caption{{\it Inverse voltage dependence of the FWS-Andreev resonances:} The
  figure shows the inverse voltage dependence of $E_{k,c}/eV$, the
  Floquet-Lippmann-Schwinger energy parameter divided by
  voltage. Panel a corresponds to $\Gamma/\Delta=0.1$, and panel b to
  $\Gamma/\Delta=0.3$.
\label{fig:spectres}
}
\end{figure}

Now that one of the Rabi resonances has been confirmed, further
numerical evidence is provided for the fully replicated Floquet
spectrum, and, more precisely, for the double FWS-Andreev ladder. A
first possibility for visualizing the two FWS-Andreev ladders (see
Fig.~\ref{fig:5}) is to plot the spectral current as a function of
energy.  This quantity measures the contribution to the total current
of FWS states at a given energy $\omega$ so that
the dc current is obtained as the integral over $\omega$
of the spectral current. Fig.~\ref{fig:5} reveals sharp resonances
with alternating signs, corresponding to the two alternating
FWS-Andreev ladders carrying opposite currents, as for Andreev bound
states at equilibrium. Fig.~\ref{fig:5} confirms the generalization to
nonequilibrium of Andreev bound states carrying opposite currents. The
insert of Fig.~\ref{fig:5} enlarges a selected resonance, which
reveals the smallness of its width in energy, due to the weakness of
the equilibrating mechanisms. The data-points shown in
Fig.~\ref{fig:5} correspond to the raw ones, evaluated with an
adaptative algorithm for integration over energy $\omega$ (the data
for Fig.~\ref{fig:5} are part of those for Fig.~\ref{fig:4}b).

The voltage dependence of the FWS-Andreev resonance energies [see
  Eqs.~(\ref{eq:1}) and~(\ref{eq:2})] is discussed now. Recursive
calculations are implemented to calculate numerically the
Floquet-Lippmann-Schwinger wave-function (see Sec. V in Supplemental
Material). Technically, the FWS-Andreev ladders are obtained from the
maxima over the energy $E_{k,c}$ of the maximum over the auxiliary
variable $l=(N_a-N_b)/2$ of the Floquet-Lippmann-Schwinger
wave-function. Eqs.~(\ref{eq:1}) and~(\ref{eq:2}) can be written as
\begin{eqnarray}
  \label{eq:3}
  \frac{E_j}{eV}&=&2j+\frac{\langle E \rangle}{\Delta}
  \times \frac{\Delta}{eV}\\
  \frac{E'_{j'}}{eV}&=&2j'-\frac{\langle E \rangle}{\Delta}
  \times \frac{\Delta}{eV}
  \label{eq:4}
  .
\end{eqnarray}
Fig. ~\ref{fig:spectres} shows plots inspired by
Ref.~\onlinecite{REF14}, in which the $x$-axis is $\Delta/eV$ and the
$y$-axis is $E_{k,c} /eV$ ($E_{k,c}$ is the energy of the
quasiparticle injected at wave-vector $k$ into lead
$S_c$). Fig.~\ref{fig:spectres}a (for $\Gamma/\Delta=0.1$) and
Fig.~\ref{fig:spectres}b (for $\Gamma/\Delta=0.3$) reveal two ladders
of Andreev resonances compatible with Eqs.~(\ref{eq:3})
and~(\ref{eq:4}). A ladder with positive slope $\langle E
\rangle/\Delta$ alternates with the other one with negative slope
$-\langle E \rangle/\Delta$. In addition, numerical evidence for
repulsion between FWS-Andreev resonances is obtained at the crossing
points, due to Landau-Zener-St\"uckelberg transitions. A qualitative
agreement is obtained in the voltage-dependence of the normalized
current $hI_c/e\Delta$, and of the spectrum of FWS-Andreev resonances
(obtained from the Floquet wave-function; see
Fig.~\ref{fig:spectres}). Namely, the current is controlled by
$eV_R/\Delta$ slightly above $eV_R/\Delta\simeq \Delta/10$ while the
avoided crossings are at $\Delta/eV_1\simeq 14$. Quantitatively, the
voltage $V_1$ is clearly smaller than $V_R$. A first explanation is
that the current~$I_c$ does not depend only  on the resonance spectrum
but also on the full FWS-Andreev
wave-function and on the occupation of the FWS-Andreev resonances. On
the other hand, more trivially, no obvious reason can be advocated of
why perfect quantitative agreement is expected on cross-over values of
different quantities calculated in different manners. Further
investigations involve recalculating the current $I_c$ with
FWS-Andreev wave-functions, but this goes beyond the scope of the
present paper.

The paper deals solely with the current $I_c$ through lead $S_c$ (see
Fig.~\ref{fig:1}), which corresponds to a pair current due to the
quartet or multipair mechanism, with vanishingly small quasiparticle
component due to MARs. All contacts between the quantum dot and the
superconducting leads have identical transparency and, together with
the assumption of particle-hole symmetry, it can be demonstrated
\cite{REF10} that the corresponding $I_c$ is due solely to
correlations among Cooper pairs, without quasiparticles. It is thus
acceptable to address a link between equilibrium and nonequilibrium
DC-superflows, without the contribution of MARs in the current, except
for the contribution of those in the line-width broadenings.

\begin{figure}[htb]
  \includegraphics[width=.45\textwidth]{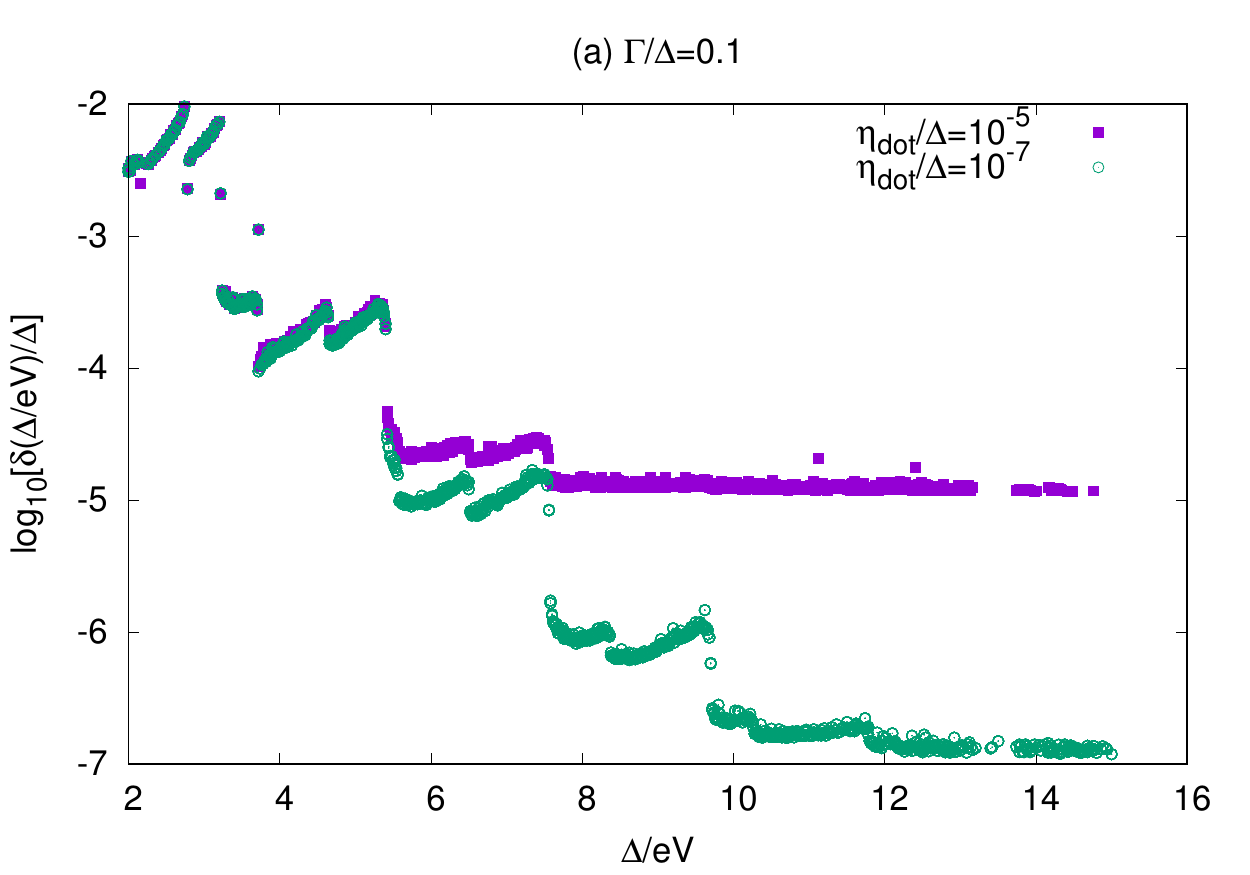}

  \includegraphics[width=.45\textwidth]{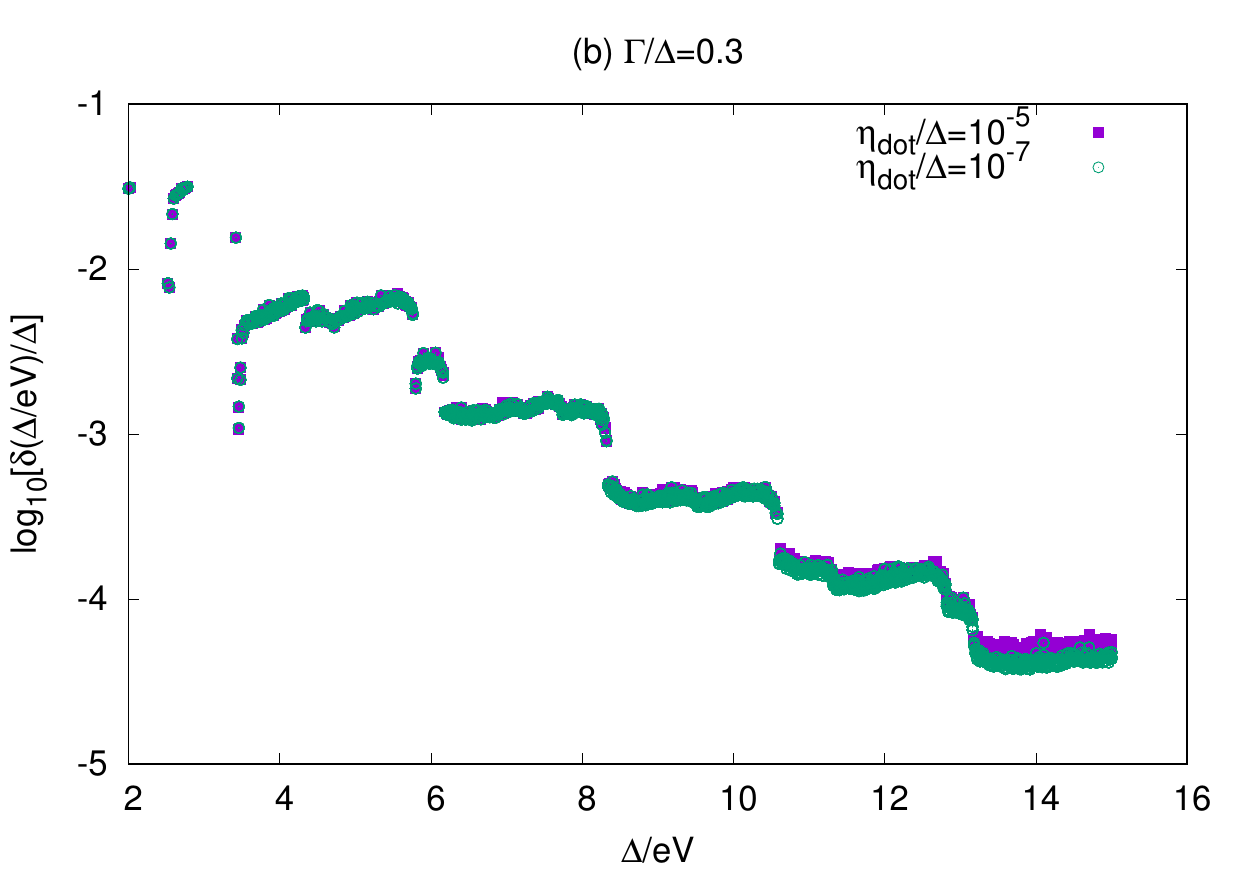}
\caption{{\it Inverse voltage dependence of the line-width
    broadening:} The figure shows the variations of the logarithm of
  the normalized line-width broadening $\delta/\Delta$ of the
  FWS-Andreev resonances as a function of the normalized inverse
  voltage $\Delta/eV$, for the values of $\eta_{dot}/\Delta$ shown on
  the figure. Panel a corresponds to $\Gamma/\Delta=0.1$ and panel b
  to $\Gamma/\Delta=0.3$.
\label{fig:6}
}
\end{figure}

As discussed above, the equilibrating coupling to the semi-infinite
quasiparticle continua is exponentially small in the inverse of
voltage: In the FWS-Andreev viewpoint, this coupling is due to
tunneling from the FWS-Andreev ladders to the continua, through a
classically forbidden region of length inverse proportional to the
bias voltage (see Sec.~\ref{sec:II}). On the other hand, tunneling
through the classically forbidden region proceeds from MARs. Each
elementary electron-hole or hole-electron process increases or reduces
by unity the value $l=(N_a-N_b)/2$ of the auxiliary variable
introduced in the preceding Sec.~\ref{sec:II} and in Sec.~VI of
Supplemental Material.  Thus, the FWS-Andreev viewpoint suggests the
appearance of step-like variations for the logarithm $\delta/\Delta$,
the normalized line-width broadening of FWS-Andreev resonances as a
function of inverse bias voltage. The voltage-dependence of
$\delta/\Delta$ is thus expected to share similarities with that of
the DC-current of MARs in a superconducting quantum point contact
\cite{REF15,REF16}.

\begin{figure}[htb]
\includegraphics[width=.45\textwidth]{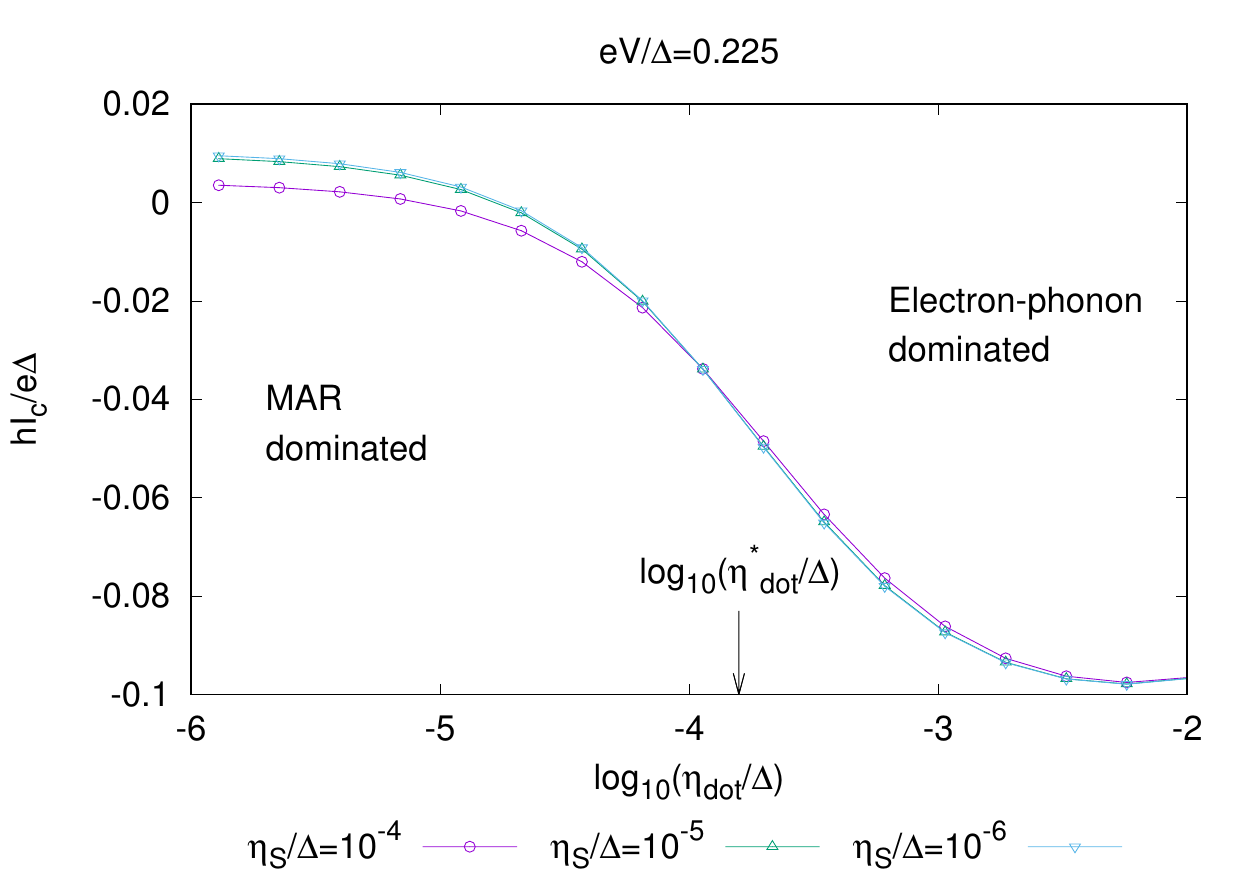}
\caption{{\it Illustration of the two regimes:} The figure shows the
  $\log(\eta_{dot}/\Delta)$-dependence of the normalized current
  $hI_c/e\Delta$, for the set-up in Fig.~\ref{fig:1}, and for the
  values of $\eta_S/\Delta$ shown on the figure, with
  $\Gamma/\Delta=0.1$ and with a generic finite value for the quartet
  phase. The data-points with $\eta_S/\Delta=10^{-5}$ almost match
  those for $\eta_S/\Delta=10^{-6}$, which is evidence for convergence
  towards adiabatic limit as $\eta_{dot}/\Delta$ and $\eta_S/\Delta$
  are reduced. The variable $\eta_{dot}^*$ is defined as the value of
  $\eta_{dot}$ at which the inflection point appears on those
  curves. The characteristic cross-over values $\hbar/\eta_{dot}^*$
  and $\hbar/\eta_S^*$ receive the interpretation of characteristic
  times for equilibration with the semi-infinite quasiparticle
  continua above the gaps via MARs. If $\eta_S=0^+$, the
  characteristic cross-over value $\eta_{dot}^*$ separates between two
  regimes: MAR-dominated equilibration for $\eta_{dot}\ll
  \eta_{dot}^*$, and electron-phonon-dominated for $\eta_{dot}\gg
  \eta_{dot}^*$. The experimental values for $\hbar/\eta_{dot}$ and
  $\hbar/\eta_S$ in a given sample are to be compared with the
  characteristic times $\hbar/\eta_{dot}^*$ and $\hbar/\eta_S^*$ in
  order to determine the dominant equilibration mechanism.
\label{fig:INTRO1}
}
\end{figure}

Those expectations for the normalized FWS-Andreev line-width
broadening $\delta/\Delta$ are fully confirmed by the numerical
results shown in Fig.~\ref{fig:6}a (for $\Gamma/\Delta=0.1$) and
Fig.~\ref{fig:6}b (for $\Gamma/\Delta=0.3$). The values of
$\delta(\Delta/eV)/\Delta$ are plotted in Fig.~\ref{fig:6}a and in
Fig.~\ref{fig:6}b as a function of inverse normalized voltage
$\Delta/eV$. The overall behavior for the envelope of
$\delta(\Delta/eV)/\Delta$ is the following:
$\log(\delta(\Delta/eV)/\Delta) \sim -\Delta/eV$, and distinguishing
features of regularly spaced steps delimited by MAR thresholds are
clearly identified in the variations of $\log[\delta(\Delta/eV)]$. The
corresponding voltages are $eV_n=\Delta/2n$, with $n=3,4,5$. The
comparison between $\eta_{dot}/\Delta=10^{-5}$ and
$\eta_{dot}/\Delta=10^{-7}$ (see Figs.~\ref{fig:6}a and b) shows that,
for $\Gamma/\Delta=0.1$, the low-voltage (large $\Delta/eV$)
normalized line-width broadening $\delta(\Delta/eV)/\Delta$ is
strongly sensitive on relaxation, through the normalized Dynes
parameter $\eta_{dot}/\Delta$. The dominant contribution to
$\delta/\Delta$ originates from $\eta_{dot}/\Delta$ at low $eV/\Delta$
and $\Gamma/\Delta$, which is in agreement with the forthcoming
discussion on Fig.~\ref{fig:INTRO1}. The sensitivity on
$\eta_{dot}/\Delta$ is much weaker for the larger value
$\Gamma/\Delta=0.3$, because the coupling to the quasiparticle
continua is much larger for $\Gamma/\Delta=0.3$ than for
$\Gamma/\Delta=0.1$. The FWS-Andreev avoided crossings in
Figs.~\ref{fig:spectres}a and b do not correlate at all with the steps
in Figs.~\ref{fig:6}a and b. This shows that, as it is discussed
above, it is MARs (and not the avoided crossings in themselves) which
are at the origin of the steps in $\log(\delta(\Delta/eV))$.

\begin{figure}[htb]
\includegraphics[width=.45\textwidth]{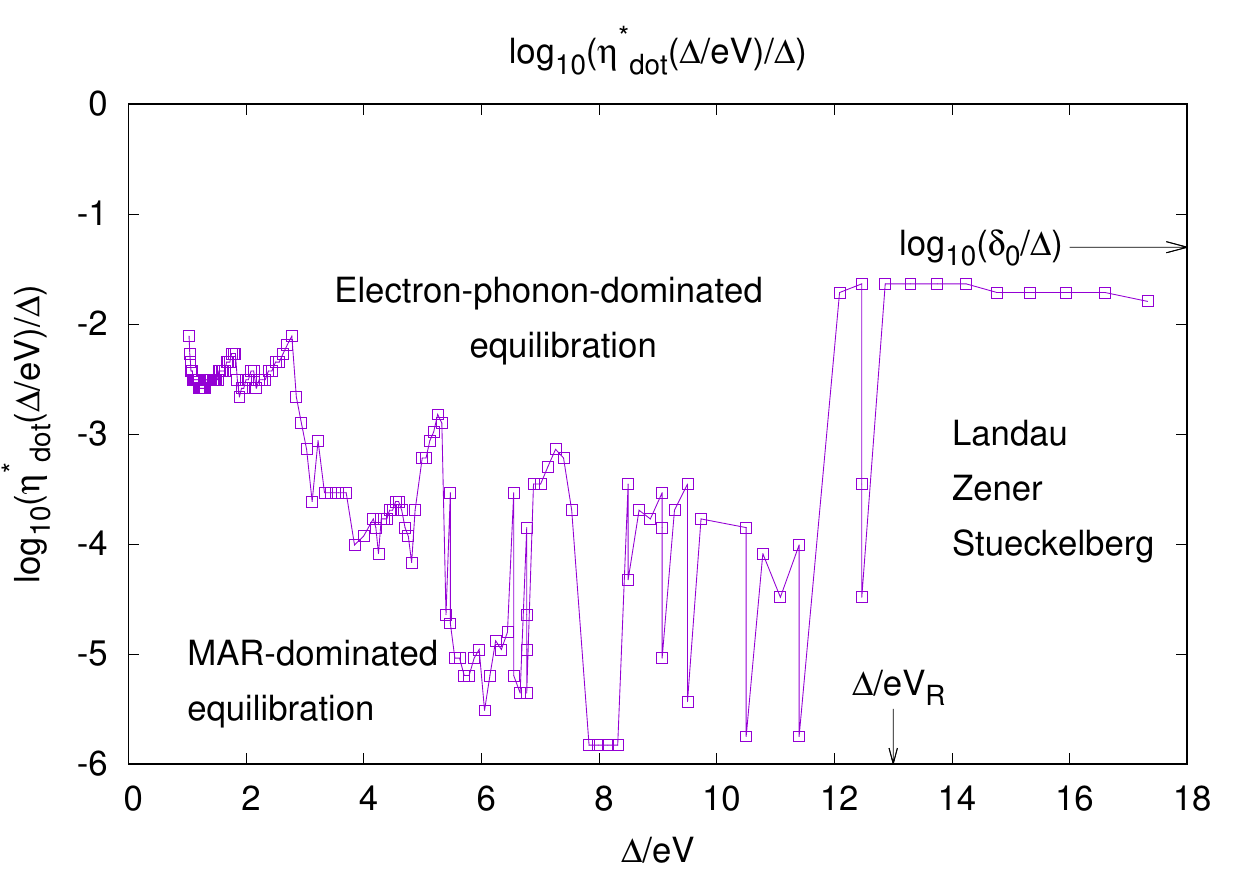}
\caption{{\it Inverse voltage dependence of $\eta_{dot}^*/\Delta$:}
  The figure shows $\eta_{dot}^*/\Delta$ as a function of $\Delta/eV$
  for the set-up in Fig.~\ref{fig:1}, and for $\Gamma/\Delta=0.1$. The
  value of $\eta_{dot}^*/\Delta$ is determined from the inflection
  points of $I_c(\eta_{dot} /\Delta)$ for all values of voltage, for
  the same data as in Fig.~\ref{fig:4}b. The arrows on the $x$ axis
  indicates $V_R$ [see Eq.~(\ref{eq:Vk}]. The arrow on the $y$ axis
  indicates $\delta_0$, the splitting between FWS-Andreev ladders at
  the avoided crossing (see Fig.~\ref{fig:spectres}).
\label{fig:7}
}
\end{figure}

Fig.~\ref{fig:INTRO1} shows for the set-up in Fig.~\ref{fig:1} the
variations of the normalized current $hI_c/e\Delta$ through lead $S_c$
at fixed voltage $V$, as a function of $\log(\eta_{dot}/\Delta)$. The
different curves correspond to Dynes parameters
$\eta_S/\Delta=10^{-6}, 10^{-5}, 10^{-4}$. The current becomes almost
independent on $\eta_{dot}/\Delta$ and $\eta_S/\Delta$ if those Dynes
parameter ratios are sufficiently small. In this regime, the current
takes its adiabatic limit value (corresponding to
$\eta_{dot}/\Delta=0^+$ and $\eta_S/\Delta=0^+$; see the discussion in
Secs.~I and II of Supplemental Material). Considering for simplicity
the limit $\eta_S=0^+$, the characteristic cross-over value
$\eta_{dot}^*/\Delta$ of the Dynes parameter $\eta_{dot}/\Delta$ is
defined practically as the inflection point in the variations of
$I_c(\log(\eta_{dot}/\Delta))$ (see Fig.~\ref{fig:INTRO1}). The
equilibration time $\hbar/\eta_{dot}^*$ corresponds to the
characteristic time for connecting by MARs the low-energy quantum dot
degrees of freedom to the semi-infinite quasiparticle continua above
the gaps. Crossing-over between MAR-dominated equilibration
($\eta_{dot}\ll\eta_{dot}^*$) and electron-phonon-dominated
equilibration ($\eta_{dot}\gg\eta_{dot}^*$) changes the sign of $I_c$
in Fig.~\ref{fig:INTRO1}. The nature of the dominant equilibrating
mechanism can thus produce important qualitative changes in the value
of physical observables such as the current $I_c$. Thus, we discuss
here the possibility of important deviations with respect to our
previous work \cite{REF10}, in which $\eta_{dot}=\eta_S=0$. The
smallness of $\eta_{dot}^*/\Delta$ makes it plausible that, depending
on the experimental values for the Dynes parameters and for bias
voltages, the regime $\eta_{dot}\gg\eta_{dot}^*$ can be more relevant to
experiments \cite{REF4} than the limit $\eta_{dot}=0$ considered
previously \cite{REF10}. A planned project related to a comparison of
this theory with the Weizmann group experimental data \cite{REF4} is
mentioned in the concluding Sec.~\ref{sec:planned-study}.

The normalized inverse voltage-$\Delta/eV$ dependence of
$\eta_{dot}^*$ is presented in Fig.~\ref{fig:7}, in the limit
$\eta_S=0^+$. The characteristic cross-over value
$\eta_{dot}^*/\Delta$ is determined from the inflection point in
$I_c(\log(\eta_{dot}/\Delta))$ (see Fig.~\ref{fig:INTRO1}). The change
at $\Delta/eV\simeq 12$ is in between $\Delta/eV_R\simeq 10$ [$V_R$ is
the voltage associated to the resonance seen on Fig.~\ref{fig:4}]
and $\Delta/eV_1\simeq 14$ [$V_1$ is the voltage at which the avoided crossing occurs in
Fig.~\ref{fig:spectres} and defined in Eq.~(\ref{eq:Vk})]. 
Considering first the high-voltage regime
($\Delta / eV \alt \Delta / eV_R$), the overall voltage dependence of
the envelope of $\log(\eta_{dot}^*/\Delta)$ is compatible with the
scaling $\log(\eta_{dot}^*/\Delta)\sim -\Delta/eV$ (see
Fig.~\ref{fig:7}), which coindices with the same overall scaling for
the spectral quantity $\log(\delta/\Delta)$ (see
Fig.~\ref{fig:6}a). The nonmonotonous behavior of the voltage
dependence of $\eta_{dot}^*/\Delta$ (see Fig.~\ref{fig:7}) is
compatible with a strong enhancement of $\log(\eta_{dot}^*/\Delta)$ at
the MAR thresholds, such as the first terms in the series
$eV_n=\Delta/(2n+1)$, with $n=1,2,3$. It is expected that coupling the
quantum dot by MARs to the gap edge singularities of one of three
superconductors is like resonantly coupling the quantum dot to a
normal bath, producing strong equilibration at the specific voltages
of the MAR thresholds. On the other hand, the value of
$\log(\eta_{dot}^*/\Delta)$ is much larger if $\Delta/eV \agt
\Delta/eV_R$, which matches the onset of coherence between the two
FWS-Andreev ladders. The value of $\eta_{dot}^*$ for $\Delta/eV \agt
\Delta/eV_R$ is compatible with $\delta_0$, the splitting between
FWS-Andreev resonances (see Fig.~\ref{fig:spectres}). It was mentioned
above that the interpretation of Fig.~\ref{fig:4} (regarding the
comparison between $\eta_{dot}/\Delta=0$ and
$\eta_{dot}/\Delta=10^{-3}$) implies that very low characteristic
scales of $\eta_{dot}/\Delta$ appear in $I_c(\log(\eta_{dot}/\Delta))$
at low bias voltage, even if $V\alt V_R$. This implies multiple values
for $\eta_{dot}^*/\Delta$ if $V\alt V_R$, which are hidden from
Fig.~\ref{fig:7} in order to keep the presentation of this cross-over
diagram as simple as possible. Obvious difficulties arise in the
numerics to detect with an adaptative integration algorithm all of the
FWS-Andreev resonances contributing significantly to the spectral
current at low $V/\Delta$ and very low-$\eta_{dot}/\Delta$, because
the width of those resonance is tiny in this parameter range. The raw
data (which are not presented here) for the variations of
$I_c(\log(\eta_{dot}/\Delta))$ with $\log(\eta_{dot}/\Delta)$ were
carefully inspected in voltage window $V\alt V_R$. Inflection points
appear in $I_c(\log(\eta_{dot}/\Delta))$ at very low
$\eta_{dot}/\Delta$ (the detected inflection points are typically in
the interval $-6\alt \log(\eta_{dot}/\Delta)\alt -4.5$.). An overall
trend to complex variations of $I_c(\log(\eta_{dot}/\Delta,eV/\Delta)$
is obtained in addition in the same voltage window. Qualitatively, the
behavior of $\eta^*_{dot}/\Delta$ correlates with spectral properties
such as the FWS-Andreev line-width broadening $\delta$ due to the
coupling to the semi-infinite quasiparticle continua of the
FWS-Andreev ladders of resonances (see Fig.~\ref{fig:6}), because both
quantities are exponentially small in $\Delta/eV$. However,
quantitatively, it is not possible to determine to which extent the
onset of the Landau-Zener-St\"uckelberg region of the cross-over
diagram upon reducing $V/\Delta$ is closer to $V_R$ than to $V_1$. The
conclusion is thus that it is remarkable that qualitatively, the
low-energy scales in the current relate to those in FWS-Andreev
spectrum. However, the correspondence is not quantitative, for the
obvious reasons which were already mentioned above.

\section{Conclusions and discussion}
\label{sec:V}
\subsection{Conclusions}
At equilibrium, the supercurrent is calculated solely in terms of to
the spectrum of Andreev bound states once the states at negative
energy have received their equilibrium occupation numbers at zero
temperature. A less demanding relation between the current $I_c$
through lead $S_c$ and spectral properties was addressed in the more
complex situation of relaxation due to the generalized Dynes parameter
$\eta_{dot}$ (encoding relaxation due to the coupling to phonons), and
of a coupling by MARs to the semi-infinite quasiparticle
continua. More precisely, it was demonstrated numerically that,
qualitatively, the characteristic low-energy scales in the
supercurrent $I_c$ through lead $S_c$ are related to those in the
FWS-Andreev spectrum of resonances, because those have similar
voltage-dependence. Three low-energy (long-time) scales emerge in
those numerical calculations based on Keldysh Green's functions and on
the Floquet-Lippmann-Schwinger dressing algorithm: First, already for
a single FWS band, the spectral line-width broadening $\delta$ of FWS
resonances is exponentially small in $\Delta/eV$ (with $\Delta$ the
superconducting gap), due to the coupling by multiple Andreev
reflections to the semi-infinite quasiparticle continua above the
gaps. Second, the degeneracy of the FWS-Andreev two-level system is
lifted by an amount $\delta_0$ at avoided crossings. Third, the
cross-over value $\eta^*$ of the Dynes parameter $\eta$ is related to
the speed of convergence of the current towards adiabaticity in the
limit $\eta\rightarrow 0^+$.

The qualitative correspondence between transport
and spectral properties is however not quantitative, which is maybe
not unexpected. In addition, the current $I_c$ also couples to the
FWS-Andreev wave-functions and to the populations of the FWS-Andreev
ladders of resonances. Given the complexity of the set-up, it is
remarkable that simple ideas such as a weak version of this relation
between transport and the spectrum can be used to understand
qualitatively the numerical calculations of the characteristic
low-energy/long time scales in the current $I_c$. It is within the
achievements of the Keldysh dressing algorithm to obtain exact
numerical values for the currents in the set-up of interest
here. However, it is not within the scope of the Keldysh algorithm to
provide by itself simple intuitive physical pictures. A full solution
of the FWS-Andreev viewpoint (including calculation of the
wave-functions and populations) is more promising to obtain simple
physical explanations. The present paper is a first step along this
roadmap.

Cavity quantum electrodynamics experiments are promising to perform
the spectroscopy of the FWS-Andreev ladders of resonances (see for
instance the recent Ref.~\onlinecite{Bruhat}). At present time, no
obvious reason can be advocated on why the nonequilibrium FWS-Andreev
two-level system should show experimental relaxation times much larger
than that of the equilibrium Andreev two-level system
\cite{quantronics1,quantronics2}. However, further studies are
required in order to determine whether quasiparticle poisoning can be
reduced by the applied bias voltage. At present stage, the best is
probably to make the experiment without prejudice.

It was found that the normalized voltage-$V/\Delta$ dependence of the
normalized line-width broadening $\delta(eV/\Delta)/\Delta$ of
FWS-Andreev resonances is exponentially small in $\Delta/eV$, which is
the expected behavior: in the FWS-Andreev viewpoint, tunneling between
the continua and FWS-Andreev ladders proceeds through a classically
forbidden region of length proportional to $\Delta/eV$. In addition,
the normalized voltage-$V/\Delta$ dependence of the cross-over
normalized Dynes parameters $\eta_{dot}^*(V/\Delta)/\Delta$ was
evaluated, which is in the first place a quantity of physical
interest. In the absence of coupling between the two FWS-Andreev
ladders at voltage $V\agt V_R$, the cross-over normalized Dynes
parameter $\eta_{dot}^*(V/\Delta)/\Delta$ corresponds to the inverse
of the characteristic time scale for connecting by MARs the low-energy
quantum dot degrees of freedom with the quasiparticle semi-infinite
continua above the gaps. Thus, $\hbar/\eta_{dot}^*$ is the
characteristic time scale for equilibration with those quasiparticle
continua if $V\agt V_R$. The characteristic time $\hbar/\eta_{dot}^*$
is of order $\hbar/\eta^*_{dot} \sim 1/\delta_0$ if $V\alt V_R$, where
$\delta_0$ is the splitting of the FWS-Andreev two-level system. In
addition, $\eta_{dot}^*$ is multiply defined if $eV\alt V_R$, and much
lower characteristic $\eta_{dot}^*$ emerge for $V\alt V_R$, with
complex behavior of the related $I_c(\eta_{dot}/\Delta,V/\Delta)$.

\subsection{Perspectives on the recent Weizmann group experiment \cite{REF4}}
\label{sec:planned-study}
The interpretation of the Weizmann group cross-correlation experiment
\cite{REF4} is the following: A train of quartets transmitted from
$S_c$ to ($S_a$, $S_b$) is produced, followed by another train from
($S_a$, $S_b$) to $S_c$, and so on.  This is reminiscent of thermal
effects on the noise of a two-terminal point contact at equilibrium
but there, it is the quantum coherent Landau-Zener-St\"uckelberg (LZS)
transitions (instead of incoherent thermally activated processes)
which change randomly the direction of the quartet superflow. In the
experimental Ref.~\onlinecite{REF4}, temperature is sufficiently low
that the thermally activated processes are negligibly small in the
noise cross-correlations (those are exponentially suppressed in the
ratio between the BCS gap and the temperature). A noticeable
difference between LZS transitions and thermal activation lies in full
quantum coherence for the former, with possible correlations among
huge numbers of Cooper pairs. In addition, in realistic situations, a
temperature dependence of the generalized Dynes parameter could
appear, as it is expected for inelastic effects. The line-width
broadening due to the continua is exponentially small in the inverse
of voltage, and the generalized Dynes parameter is exponentially small
in the inverse of temperature. A cross-over temperature comparable to
voltage is thus expected.

The present results and those of the Weizmann group \cite{REF4} share
intriguing similarities: The time scale $\sim\hbar/\delta$ (the
inverse of the FWS-Andreev line-width broadening) is possibly related
to the overall coherence time for the absolute value of the current in
this picture of LZS transitions. The scale $\sim\hbar/\delta_0$ (the
inverse of the level degeneracy at a FWS-Andreev avoided crossing) is
possibly related to the much shorter coherence time for the sign of
the current. The energy scale $\eta^*$ controls whether equilibration
is due to the coupling to the quasiparticle continua (for $\eta\alt
\eta^*$) or to inelastic electron-phonon scattering (for $\eta\agt
\eta^*$). Further calculations for the voltage and quartet phase
sensitivity of the cross-correlations are required to determine the
nature of the dominant equilibrating mechanism in the Weizmann group
experiment \cite{REF4}. In addition, gain of realism can be obtained
by discussing finite temperature effect on the quasiparticle
populations, and by treating double (instead of single) quantum dots.

The recent experimental results of the Weizmann group \cite{REF4} were
compared in the same preprint to Keldysh Green's function calculations
for the noise. The calculated current cross-correlations \cite{REF4}
provide evidence for resonance in DC-current and cross-correlations as
the quartet phase $\varphi_Q = \varphi_a + \varphi_b-2\varphi_c$ is
varied at fixed voltage $V\equiv V_a=-V_b$. This resonance
parameterized by $\varphi_Q$ is strongly sensitive on the value of the
Dynes parameter ratio $\eta_S/\Delta$: the cross-correlation signal
varies by one order of magnitude as $\eta_S/\Delta$ is changed in the
range $\eta_S/\Delta = 10^{-6}\div 10^{-3}$. This ultra-sensitivity of
current cross-correlations on small values of the Dynes parameters is
puzzling, especially with respect to what has been found here
regarding the behavior of the current as a function of
$\eta_{dot}$. This is why it would be also of interest to introduce
$\eta_{dot}$ in cross-correlation calculations, and to calculate the
cross-correlations for $\eta_{dot}/\Delta$ smaller or larger that
$\eta_{dot}^*/\Delta$. The first question to be asked is probably
whether the same inverse time scale $\hbar/\eta_{dot}^*$ controls the
current and the noise cross-correlations. Comparing with the
experimental data of the Weizmann group \cite{REF4} may then provide
useful information on the issue of whether the quantum dot degrees of
freedom are equilibrated with the quasiparticle continua in this
experiment. This may be all the more interesting in view of the
perspective on thermodynamics discussed now.

\subsection{Perspective on quantum thermodynamics}
\label{sec:concluC}
Final remarks are presented now, in connection with a perspective on
quantum thermodynamics in three-terminal Josephson junctions at the
quartet resonance. A limit can be worked out analytically in which
quasiparticles on the quantum dot never reach equilibration: the
infinite gap limit. Coupling solely to superconducting condensates
does not allow for propagation of entropy (because, in the infinite
gap limit, each superconducting lead has the condensate as a single
state, and a vanishingly small entropy). A small energy scale is
expected in the presence of a finite gap and small $\eta_S/\Delta$ or
$\eta_{dot}/\Delta$, related to the interplay with the finite
line-width broadening of FWS-Andreev resonances due to the coupling to
quasiparticle continua above the gaps. Future evaluations of the heat
current in a three-terminal Josephson junction biased in the quartet
voltage configuration can thus bring informations complementary to
those presented above.

To conclude, those three-terminal Josephson junctions can be viewed as
prototypical examples of ``half-open systems''.  Those systems share
features of open systems (for example: extended quasiparticle states
above the gaps, finite current due to bias voltages on the leads) with
the apparently contradicting features of closed system (for example:
strong confinement of the quantum dot degrees of freedom, poor
equilibration, sharp Andreev resonances in the Floquet spectrum).

\section*{Acknowledgments}
The authors acknowledge support from ANR Nanoquartets 12-BS-10-007-04.
The authors thank the CRIANN in Rouen for use of its computing
facilities. Part of the numerical calculations were also performed on
the local computing facilities of Institut N\'eel in
Grenoble. R.M. acknowledges fruitful discussions with Denis Feinberg,
especially on the infinite-gap limit. R.M. thanks Jean Christian
Angl\`es d'Auriac for having provided useful advice with daily
numerics. R.M. and B.D. thank Didier Mayou for useful remarks on
phonons in connection with Ref.~\onlinecite{Mayou}. R.M. acknowledges
stimulating discussions on thermodynamics with Alexia Aff\`eves and
G\'eraldine Haack. R.M. thanks the ``Fondation Nanosciences'' in
Grenoble for financial support and useful discussions in the framework
of Yuli Nazarov's Chair of Excellence. R.M. thanks his collaborators
from the Weizmann Institute: Yuval Cohen, Yonathan Ronen, Hadas
Shtrickman and Moty Heiblum. R.M. wishes to express special thanks to
Denis Basko for in-depth reading of a previous version of the
manuscript, and for useful suggestions in connection with his recent
solution \cite{Basko1} of a related time-dependent problem
\cite{Basko2}.  R.M. and B.D. thank Pascal Degiovanni for his previous
collaboration on a related topic \cite{preprint}. Finally, R.M.,
J.G.C. and B.D. thank Jean-Jacques Pr\'ejean for his warm hospitality
at the time where those ideas emerged, and for useful comments on the
manuscript.

\appendix

\section*{The Hamiltonian}
The Hamiltonian of a three-terminal Josephson junction (see
Fig.~\ref{fig:1}) is the following:
\begin{eqnarray}
&&\hat{\cal H}(t)=(\mbox{\ref{eq:A1}}) +
(\mbox{\ref{eq:A2}}) + (\mbox{\ref{eq:A3}})(t)\\
\label{eq:A1}
&&(\mbox{\ref{eq:A1}})= \sum_{j,{\bf k},\sigma} \epsilon_{\bf k} c^+_{j,{\bf k},\sigma}
c_{j,{\bf k},\sigma} \\
\label{eq:A2}
&&(\mbox{\ref{eq:A2}})=\sum_{j,{\bf k}} \left[\Delta_j c_{j,{\bf
      k},\uparrow}^+ c_{j,-{\bf k},\downarrow}^+ + \Delta_j^*
  c_{j,-{\bf k},\downarrow} c_{j,{\bf k},\uparrow} \right]\\ \nonumber
&&(\mbox{\ref{eq:A3}})(t)=\\
\label{eq:A3}
&&\sum_{j,{\bf k},\sigma} J_{j,{\bf k}} \left( e^{-is_j\omega_0 t}
c_{j,{\bf k},\sigma} d_{\sigma} +e^{is_j\omega_0 t} d_\sigma^+
c_{j,{\bf k},\sigma}\right) .
\end{eqnarray}
The first term (\ref{eq:A1}) corresponds to the kinetic energy in each
of the superconducting leads labels by $j\in\{S_a , S_b , S_c \}$. The
second term (\ref{eq:A2}) is the mean-field BCS pairing in lead $j$,
with $\Delta_j = |\Delta_j| \exp(i \varphi_j)$. The gauge is chosen in
such a way as the phase $\varphi_j$ is time-independent, and the time
$t$-dependence is in the tunnel term (\ref{eq:A3})$(t)$. The
Hamiltonian is time-periodic and, in the quartet configuration of bias
voltages, we have $s_j\in\{0,\pm1\}$. It is only in Sec.~V of
Supplemental Material that the wave-vector dependence of the tunnel
terms is retained.  The calculations presented in the remaining of the
paper correspond to $J_{j,q}\equiv J_j$, and, instead of $J_j$, the
contact transparencies are parameterized by the more usual parameters
$\Gamma_j=J_j^2/W$, where $W$ is the band-width. If Dyson equations
are involved, it is then convenient to view the $J_j$ terms as a
self-energy. In this context, those $J_j$ are denoted by $\Sigma_j$
(see Supplemental Material).

\end{document}